\documentclass[a4paper,usenatbib]{mnras}

\usepackage{newtxtext}

\usepackage[T1]{fontenc}
\usepackage{ae,aecompl}

\usepackage{amsmath,amstext,amsgen,amsbsy,amsopn,amsfonts,theorem}

\usepackage[pdftex]{graphicx}
\usepackage{xspace}
\usepackage{amsmath}
\usepackage{hyperref}
\usepackage{framed}
\usepackage{txfonts}
\usepackage{epstopdf}
\usepackage{gensymb}
\usepackage{float}

\usepackage[normalem]{ulem}

\usepackage{graphicx}	
\usepackage{amsmath}	
\usepackage{amssymb}	

\AtBeginShipout{%
  \ifnum\value{page}>1 %
    \typeout{* Additional boxing of page `\thepage'}%
    \setbox\AtBeginShipoutBox=\hbox{\copy\AtBeginShipoutBox}%
  \fi
}

\newcommand{\teff}{T_\mathrm{eff}}
\newcommand{\caii}{Ca \textsc{II}}

\title[Stellar streams around the MCs in 4D]{Stellar streams around the 
Magellanic Clouds in 4D\thanks{Based on observations collected at the European 
Organisation for Astronomical Research in the Southern Hemisphere under ESO 
programme 098.B-0454(A).}}

\author[C. Navarrete et al.]{C. Navarrete$^{1,2,3}$\thanks{Contact e-mail: 
\href{cnavarre@astro.puc.cl}{cnavarre@astro.puc.cl}}, V. Belokurov$^{3,4}$, M. 
Catelan$^{1,2}$\thanks{On sabbatical leave at The Observatories of the Carnegie 
Institution for Science, 813 Santa Barbara Street, Pasadena, CA 91101, USA}, P. Jethwa$^{5}$, 
S. E. Koposov$^{3,6}$ \newauthor J.~A. Carballo-Bello$^{1,2}$, P. Jofr\'e$^{3,7}$, D. Erkal$^{8}$, S. Duffau$^{9}$ and J.~M. Corral-Santana$^{10}$ \\
$^{1}$Instituto de Astrof\'isica, Pontificia Universidad Cat\'olica de Chile, 
Av. Vicu\~na Mackenna 4860, 782-0436 Macul, Santiago, Chile \\ 
$^{2}$Millennium Institute of Astrophysics, Av. Vicu\~na Mackenna 4860, 
782-0436, Macul, Santiago, Chile \\
$^{3}$Institute of Astronomy, University of Cambridge, Madingley Road, 
Cambridge, CB3 0HA, UK \\
$^{4}$ Center for Computational Astrophysics, Flatiron Institute, 162 5th Avenue, 10010, New York, NY, USA \\
$^{5}$ European Southern Observatory, Karl-Schwarzschild-Str. 2, 85748 Garching, Germany \\
$^{6}$ McWilliams Center for Cosmology, Department of Physics, 
Carnegie Mellon University, 5000 Forbes Avenue, Pittsburgh, PA 15213, USA \\
$^{7}$ Nucleo de Astronom\'ia, Facultad de Ingenier\'ia, Universidad Diego
Portales, Av. Ejercito 441, Santiago, Chile\\
$^{8}$ Department of Physics, University of Surrey, Guildford, GU2 7XH, UK \\
$^{9}$ Universidad Andr\'es Bello, Departamento de Ciencias F\'isicas, Facultad de Ciencias Exactas, Fern\'andez Concha 700, Las Condes, Santiago, Chile \\
$^{10}$ European Southern Observatory (ESO), Alonso de C\'ordova 3107, Vitacura, Casilla 19, Santiago, Chile }

\date{Last updated 2018 xxxx xx; in original form 2018 xxxx xx}

\pubyear{2018}

\begin{document}
\label{firstpage}
\pagerange{\pageref{firstpage}--\pageref{lastpage}}
\maketitle

\begin{abstract}

We carried out a spectroscopic follow-up program of the four new
stellar stream candidates detected by \citet{Be2016} in the outskirts
of the Large Magellanic Cloud (LMC) using FORS2 (VLT). The
medium-resolution spectra were used to measure the line-of-sight
velocities, estimate stellar metallicities and to classify stars into
Blue Horizontal Branch (BHB) and Blue Straggler (BS) stars. Using the
4-D phase-space information, we attribute approximately one half of
our sample to the Magellanic Clouds, while the rest is part of the
Galactic foreground. Only two of the four stream candidates are
confirmed kinematically. While it is impossible to estimate the exact
levels of MW contamination, the phase-space distribution of the entire
sample of our Magellanic stars matches the expected velocity gradient
for the LMC halo and extends as far as 33 deg (angular separation) or
29 kpc from the LMC center. Our detections reinforce the idea that the
halo of the LMC seems to be larger than previously expected, and its
debris can be spread in the sky out to very large separations from the
LMC center. Finally, we provide some kinematic evidence that many of
the stars analysed here have likely come from the Small Magellanic
Cloud.
\end{abstract}

\begin{keywords}
Magellanic Clouds -- Galaxy: halo -- stars: horizontal branch
\end{keywords}



\section{Introduction}

The Large and Small Magellanic Clouds (LMC and SMC, respectively) are a pair of 
nearby, likely massive dwarf satellite galaxies, probably orbiting the Milky Way 
(MW). Located at Galactocentric distances of $\approx$50 and $\approx$60 kpc, 
respectively, at the moment they are well within the halo of the MW. In the 
hierarchically assembled Universe, the LMC should have accreted smaller objects 
whose tidal debris would eventually mix and dissolve in the satellite's 
gravitational potential, forming its stellar halo. Therefore, the existence of 
the LMC stellar halo and its extent and lumpiness could be used to understand 
some of the details of the structure formation on scales below 
L${}_{\ast} \sim 2 \times 10^{10} L_{\odot}$ (characteristic luminosity of MW 
and Andromeda-like galaxies).
 
The scenario in which the LMC had been accreting smaller systems to
form its own stellar halo appears to be reinforced by the discovery of
a group of ultra-faint objects in the vicinity of the Magellanic
Clouds \citep[MCs;][]{Koposov15, Bechtol15,
  DrlicaWagner15,Kim15,maglites2016,Carinas}. Using dynamical models
of the LMC in-fall, \cite{Jethwa16} found that at least one-third of
these new objects could be associated with the LMC, with the Cloud's
total dwarf population reaching as many as $\sim$70 in the
past. Note that most recently, based on the newly proper
  motion measurements from Gaia DR2 \citep{GaiaHelmi2018, Fritz2018a},
  the association to LMC of these dwarfs have been confirmed or
  disapproved, depending on the sample of member stars considered
  \citep[see,
    e.g.,][]{Simon2018,Kallivayalil2018,Fritz2018b,PaceLi2018}. Moreover,
the complex morphology of HI gas surrounding the MCs \citep[see the
  review of][and references therein]{DOnghia16}, is a living proof of
the intricate dynamical interaction history of the LMC and SMC -- both
between each other and with the MW -- of which little has been
understood to date. Only recently, based on high precision proper
motions derived using the {\it Hubble Space Telescope}, the fast
tangential motion of the LMC \citep{Kallivayalil06, Kallivayalil13}
has been uncovered, implying a large orbital velocity which in turn
favours a recent accretion onto the MW ($<$4 Gyr). This scenario has
strong implications for the mass of the Galaxy \citep{Busha11} and the
genesis of the gaseous Magellanic Stream \citep[MS;
  e.g.,][]{Besla10}. It has been suggested that close encounters
between the LMC and SMC may be responsible for both the gas and
stellar structure identified around these galaxies
\citep[e.g.][]{Besla12, Diaz12}.  Moreover, according to the
state-of-the-art simulations of the interaction between the LMC, the
SMC and the MW, large sprays of the SMC debris ought to be discovered
throughout many sightlines around the LMC \citep[see][]{Besla12,
  Besla13, Diaz11, Diaz12, Hammer15}.

Early attempts to map out the stellar halo of the LMC were based on
star count maps \citep[e.g.,][]{Irwin91, Kinman91} which later were
found to be compatible with extended disk models \citep{Alves04}, and
the kinematics of specific and rare tracers such as RR Lyrae stars
\citep{Minniti03} and planetary nebulae \citep{Feast68} supporting the
existence of an extended spheroidal component. Nonetheless, the LMC
stellar disk has been found to stretch as far as $\sim$10
scale-lengths \citep[$\sim$15 deg from the LMC centre,
  e.g.,][]{Saha10, Balbinot15}.  Therefore, to study and detect the
LMC's stellar halo, the outskirts of the galaxy need to be explored,
where the LMC's disk is not as overwhelming. For instance,
\cite{Munoz2006} and \cite{Majewski09} presented the first pieces of
evidence for an extended halo-like structure for the LMC traced with
spectroscopically confirmed giants in the direction of the Carina
dwarf spheroidal galaxy.

With the most recent releases of the wide-field photometry from the
Dark Energy Survey \citep[DES,][]{Diehl14} and the Gaia mission
\citep{Gaia16}, the low-surface brightness structure of the Clouds has
started to come into a sharp focus \citep{Be2016, Mackey16,
  Belokurov17, Deason17, Pieres2017,Mackey2018}. These studies provided plenty of
tantalizing evidence for the past and ongoing encounters between the
Clouds as well as their disruption by the MW. However, what all of
these studies have lacked so far is the kinematic dimension. Without
the velocity information, it is fatuous to believe that the details of
the Clouds' interaction can be deciphered. In this paper, we describe a
spectroscopic effort to provide the missing line-of-sight velocity
information for a large sample of stars scattered throughout the
Magellanic System. Our targets are selected from the sample of Blue
Horizontal Branch (BHB) star candidates from \cite{Be2016} and cover a
wide range of angular distances from the LMC, i.e. between 13.0 and
48.4 deg. These stars are between 13.0 and 42.0 degrees away from the
SMC.

This paper is organised as follows. Section~\ref{sec:obs} gives the
details of the follow-up spectroscopic observations, including the
spectral fitting, the separation between Blue Stragglers and the BHBs
as well as the metallicity estimation based on the Ca II K
line. In Section~\ref{sec:lmc_smc}, we discuss the possibility of the
Magellanic origin for some of the stars in our sample and compare
their distribution in the phase-space to the predictions of the
numerical simulations of the Magellanic in-fall. The summary of our
study can be found in Section~\ref{sec:dandc}.


\section{Observations}
\label{sec:obs}

BHB candidates in the four different substructures (S1-S4) identified
by \cite{Be2016} were selected for follow-up
spectroscopy. Medium-resolution ($R\simeq$ 1\,400) spectra were
collected at Paranal Observatory, using the FORS2 spectrograph mounted
on the VLT UT1 8m telescope. The data were collected during four
nights of Visitor Mode observations (Program ID 098.B-0454A) carried
out between November 1-5 2016. The exposure times varied from
  240s up to 780s, depending on the target magnitude and airmass. The
  seeing conditions were generally good, with a mean seeing of
  0.9$\arcsec$ and with a handful of exposures with somewhat inferior
  seeing ($\sim$1.5$\arcsec$). The instrument was used with the E2V
detector, binning of 2$\times$2, SR collimator and a
1.0$\arcsec$ slit (long-slit mode). The grism used was 1200B+97, with
a dispersion of 0.36 $\AA$ per pixel. This configuration provides a
wavelength range of 3600 - 5110 \AA.

The spectra of 104 targets were obtained. Of these, 25 came from
contaminating classes of objects, such as quasi-stellar objects
(QSOs), white dwarfs or hot subdwarfs (without any Balmer line). To
reduce the contamination, from the second night onwards only stars
with colours $(g-i) < 0.0$ were considered. This additional colour cut
allows us to discard most of the QSOs \citep{Deason2012}. As a result,
79 of the 104 targets observed are likely A-type stars based on the
presence of strong Balmer lines.

The data reduction was performed using the {\tt ESOREX} pipeline
provided by ESO. Bias-subtraction, flat fielding correction, spectral
extraction, sky correction and wavelength calibration were performed
through different recipes of the pipeline. Cosmic ray hits
  were removed through the optimal extraction as implemented in the
  pipeline. No further removal was required given that our exposure
  times are relatively short, under 15 min. The spectra were not flux
calibrated. Extracted spectra were normalized using a fifth order
polynomial.

\subsection{Radial velocities} \label{sec:rvs}



To fit the Balmer lines, a S\'ersic profile \citep{Sersic68} was
adopted:
\begin{equation}
 y = a \exp{\left[-\left(\frac{\left|x-x_0\right|}{b}\right)^c\right]}\text{,}
\end{equation}\noindent
where the $a$, $b$ and $c$ parameters correspond to the line depth at
the line centre, a measure of the line width, and a proxy for the line
shape, respectively \citep[see also][]{Xue2008}. The $x_0$ coefficient
corresponds to the wavelength of the line centre and it is related to
the radial velocity, $v_r$, by $x_0 = \lambda_0(1+v_r/c)$. To perform
the spectral fit, we only consider the Balmer lines from $H_{\beta}$
to $H_{\eta}$ (i.e., from $\lambda > 3800\,{\AA}$) because a reliable
continuum normalization is more difficult for lines at bluer
wavelengths. Moreover, we consider that the six Balmer line shapes,
for a given star, have the same $b$, and $c$ parameters but different
depths. Therefore, the parameters to be estimated were $(a_1, ...,
a_6, b, c, v_r)$ plus the six parameters from the normalization of the
continuum. This model was convolved with a Gaussian profile with
$\sigma_{\lambda}$ corresponding to the mean spectral standard
deviation of the instrument profile at each pixel\footnote{Available
  in one of the output tables produced by the reduction pipeline: {\tt
    spectra\_resolution\_lss.fits}.}.

\begin{figure*}
  \includegraphics[width=\textwidth]{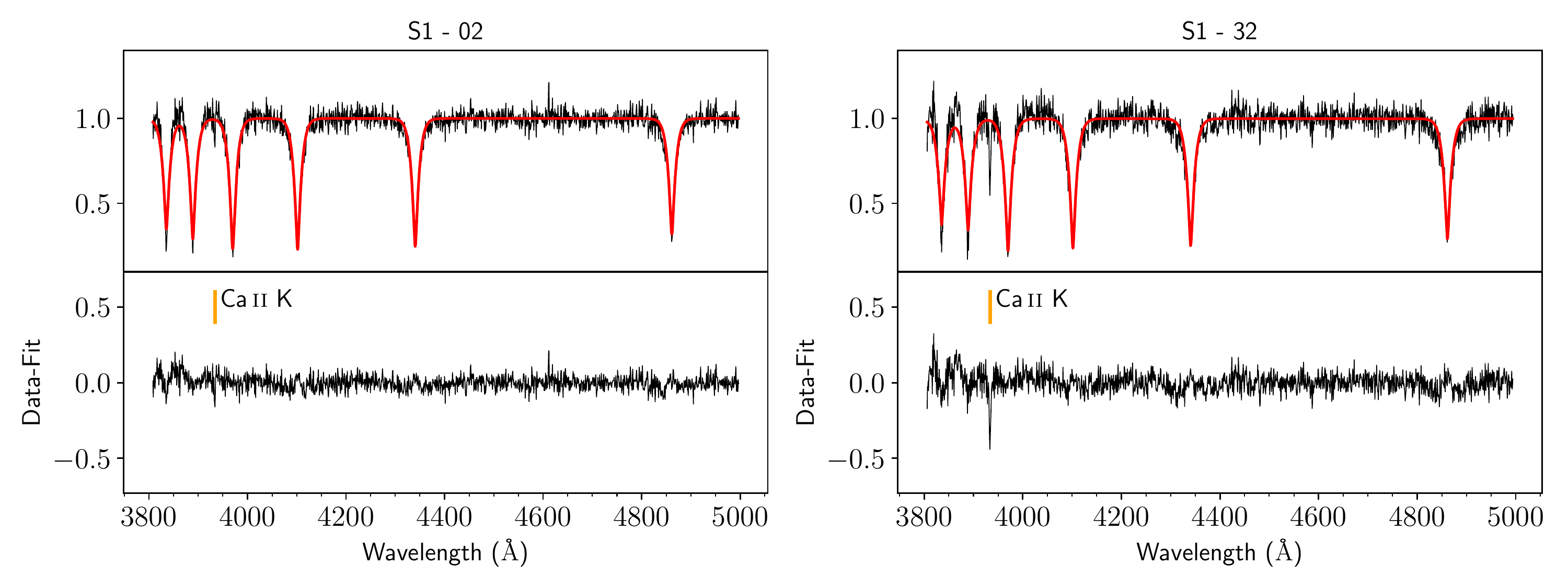}
  \caption{Two examples of reduced spectra for S1 targets. The ID of each target 
is indicated at the top of each panel. Top panels show the spectra and the 
fitting of the six Balmer lines plus the continuum normalization. Bottom panels 
show the residuals from the fit. S1~02 (left) is likely a BHB star while S1~32 
(right) is consistent with being a BS, based on the line width and shape of the 
Balmer lines (see Section~\ref{sec:bc_param}). The spectrum of S1~32 also shows 
a prominent \caii\ line at $\sim3930\, \AA$.}
    \label{fig:fit_example}
\end{figure*}

Figure~\ref{fig:fit_example} shows two examples of the
continuum-normalized spectra together with the model of the Balmer
lines (top panels) and the difference between the data and the fit
(bottom panels). In both examples, the normalization of the spectrum
tends to be less accurate at the bluest region ($\lambda <
4000\,{\AA}$), where most of the Balmer lines are located, leaving
less wavelength space for the continuum estimation. For that region,
the normalized continuum is located slightly above unity. In contrast,
a much better normalization is obtained between H$\beta$, H$\gamma$
and H$\delta$. As judged by the residuals exhibited in the bottom
panels, a S\'ersic profile provides a high fidelity description of the
Balmer lines. The most striking outlier is the \caii\ K line at
$\lambda \sim 3930\,{\AA}$; in the case of S1~32, it stands out rather
clearly, but for the star S1~02, this line is barely noticeable. In
what follows, we use the strength of the \caii\ K line to estimate
the stellar metallicity of our targets.

The 1$\sigma$ error in the fitted radial velocity parameter was
adopted as the velocity error. For most of the stars, the radial
velocity error is of the order of $\sim$5 km s$^{-1}$. Stars in the S4
substructure have the largest errors ($\sim$10 km s$^{-1}$) because
the collected spectra for these stars have slightly lower S/N than the
rest of the sample (S4 is the most distant substructure, located at
$\approx$85 kpc). Besides the error associated with the
  fitting of the Balmer lines, another source of error in the velocity
  determination is the offset in the centering of the star in the
  slit. We obtain a rough estimate of this error to be at most
  $\sim$15 km s$^{-1}$ based on the average difference in the derived
  velocities of stars observed twice and the associated error in the
  position of H$\beta$ due to a centering offset of about 0.5 pixels
  (the slit width is 4 pixels), which is the maximum offset allowed
  between the center of the star and the center of the slit before
  carrying out the exposures. This is only an upper bound on the
  actual error and thus it is not included in
  Table~\ref{tab:targets}. Note that every star would have a slightly
  different offset in the centering, which is hard to estimate in a
  star-by-star basis.

\subsection{BHB and BS separation}
\label{sec:bc_param}

While the most obvious contaminants (such as QSOs, white dwarfs, hot
subdwarfs) in our sample were discarded based on the distinct
appearance of their spectra, it is not possible (or advisable) to
distinguish between the BHBs and the Blue Straggler (BS) stars via
visual inspection. Given that \cite{Be2016} relied solely on
broad-band photometry to define a boundary between the BHBs and the
BSs, the cross-contamination of the two classes is not negligible. The
main difference between BHB and BS stars is that the former are giant
helium-burning stars while the latter are hot dwarf stars on the Main
Sequence. Therefore, it is the strength of the surface gravity that
differentiates them. Several authors \citep{Rodgers81, Kinman94,
  Wilhelm99, Clewley2002, Xue2008, Vickers12} have defined different
methods (such as $D_{0.15}$-colour or $D_{0.2}$-$f_m$ and the {\it
  scale-width-shape} methods) and the boundaries to separate BHB and
BS stars by means of their colours, in addition to Balmer line
profiles in medium-resolution spectra. The line profiles alone have
been used for this purpose as well.

Most of the previous work, however, only used $H_{\gamma}$ and/or
$H_{\delta}$ to study the Balmer line shape. The $b$ and $c$
coefficients of the S\'ersic profile are typically used to define the
scale-width-shape method to identify the BHB stars
\citep{Clewley2002}. Here, the differences in surface gravity are
reflected in the $b$ coefficient (scale width), while the effective
temperature can be measured through the $c$ (scale shape) parameter.

To take advantage of the good quality of our spectra, below we derive
a new boundary for the scale-width-shape method which makes use only
of spectroscopic measurements fitting $H_{\beta}$, $H_{\gamma}$ and
$H_{\delta}$ simultaneously.  To do so, the SDSS DR9 data \citep{A12},
from which photometry and spectra are available, were explored. We
selected A-type stars based on their spectral parameters, {$\teff$}
and $\log{(g_s)}$, and de-redenned colours $(u-g)_0$ and $(g-r)_0$, as
follows:

\begin{eqnarray}
 2.8 < \log{(g_s)} < 4.6, \nonumber \\
7500 < T_{\rm eff} < 9300\, [\textrm{K}], \nonumber \\
 0.7 < (u-g)_0 < 1.4, \nonumber \\
 -0.3 < (g-r)_0 < 0.0. 
\end{eqnarray}

Only high S/N spectra (S/N > 20) were considered and borderline cases
were excluded (i.e., 3.5 < $\log{(g_s)}$ < 4.0). Using these cuts,
$\sim$2000 A-type stars were selected, for which at least 6 Balmer
lines were visible in the wavelength range covered by the SDSS. 

\begin{figure*}
  \includegraphics[width=\textwidth]{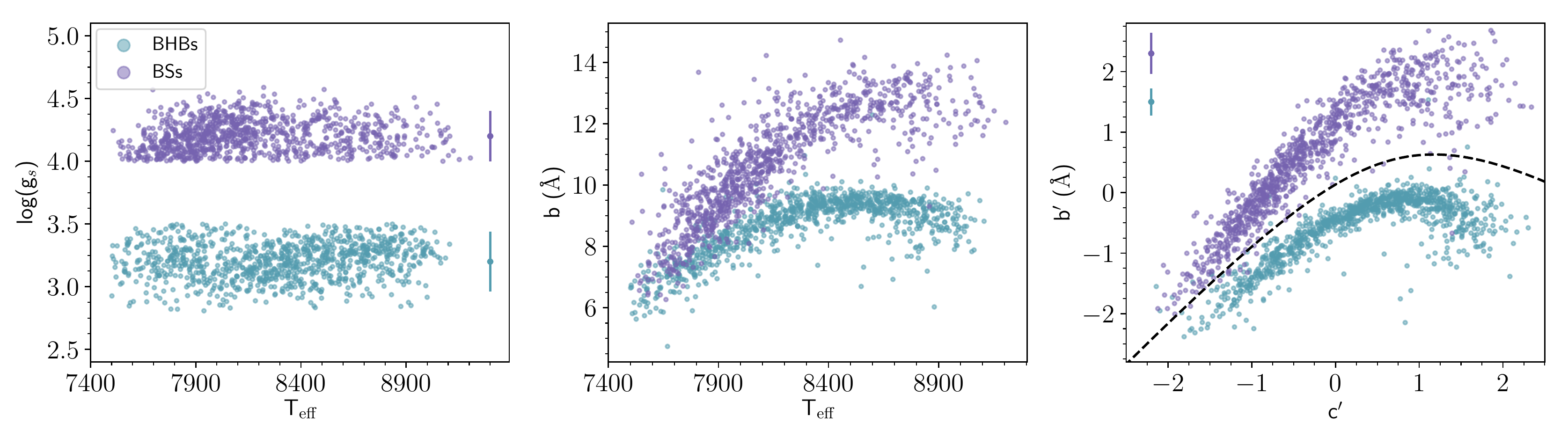}
  \caption{Spectral properties for BHB and BS stars from SDSS DR9. The
    left panel shows the effective temperature and surface gravity of
    $\simeq$ 2000 A-type stars, color-coded according to their
      most likely classification as BHB and BSs based on the surface
      gravity. The mean uncertainty in the $\log{g}$ estimate, for
      each sample, is represented by two errorbars located at the top
      left corner. The middle panel shows the $b$ parameter of the
    S\'ersic profile for the Balmer lines in those spectra as a
    function of the effective temperature. BHBs tend to have smaller
    $b$ values, i.e., narrower lines, than BSs. The right panel shows
    the $b^{\prime}$ and $c^{\prime}$ parameters of the S\'ersic
    profile (see text) and the separation between BHBs and BSs as
    derived in this work (equation~\ref{eq:border}, dashed line),
    while at the top left, the mean error of the coefficients is
    represented by two errorbars.}
    \label{fig:bhb_bs_sdss}
\end{figure*}

Following the same procedure as with our FORS2 data, a polynomial of fifth order 
plus three S\'ersic lines were fitted to the SDSS spectra, this time convolved 
with a Gaussian profile with $\sigma$ corresponding to the instrument profile 
(available for each SDSS spectrum), in the wavelength range from $\lambda 
\approx 4000$ to 5000\,{\AA}. Figure~\ref{fig:bhb_bs_sdss} shows the effective 
temperature and the surface gravity (left panel) for this SDSS sample, $T_{\rm 
eff}$ against the line width parameter $b$ (middle panel), and the two 
normalized S\'ersic parameters $b^{\prime}$ and $c^{\prime}$ (right panel), 
defined as they both have zero mean and unit variance, with 9.68, 1.70 and 0.89, 
0.11 as the mean and standard deviation of $b$ and $c$, respectively. The 
separation between BHBs and BSs is evident in all the panels, in particular for 
the stars hotter than 8300~K or $b^{\prime} >-0.5, c^{\prime} >0$. Based on the 
evident separation between BHBs and BSs in this plane, a separation boundary can 
be established. To do so, a Support Vector Machine (SVM) was used, as 
implemented in the {\tt scikit-learn} module \citep{pedregosa11}, using a linear 
kernel. The derived boundary has the form

\begin{equation} \label{eq:border}
 -0.78\,{c^{\prime}}^2 + 0.6\,{b^{\prime}}^{2} - 0.36\,b^{\prime}\,c^{\prime} + 2.05\,c^{\prime} - 2.61\,b^{\prime} + 0.34 = 0.
\end{equation}\noindent

From this equation, the boundary function B(c) was defined. Stars with
$b^{\prime}$ and $c^{\prime}$ parameter values above this boundary
division are most likely BSs, while those below are most likely
BHBs. However, those stars that are located close to the boundary can
be easily misclassified as the SDSS spectra used to derive the SVM do
not include stars with surface gravities between $3.5 < \log{(g_s)} <
4.0$, where BHBs and BSs tend to overlap.  For the training
  set, the boundary found allows a separation between BHBs and BSs
  with a completeness of BHBs of 99\% and a contamination of 18\% of
  BSs in the BHB class. These values are in stark contrast to the
  completeness and purity of photometrically-selected BHB stars:
  \cite{Vickers12} reported a 57\% completeness and 25\% contamination
  (mostly from MS A-type stars) for SDSS $(u-g)$,$(g-r)$ color-cut
  selections \citep[in good agreement with previous estimates
    by][]{Sirko2004, Bell2010}, while similar completeness and
  contamination values are found when using SDSS $(g-r)$,$(iz)$
  color-cuts (51\% and 23\%, respectively). A slightly better
  completeness sample was obtained by \cite{Fukushima18} when using
  the z-band photometry from the Hyper Suprime-Cam Subaru Strategic
  Program, with 67\% completeness and 38\% contamination.

\begin{figure*}
  \includegraphics[width=\textwidth]{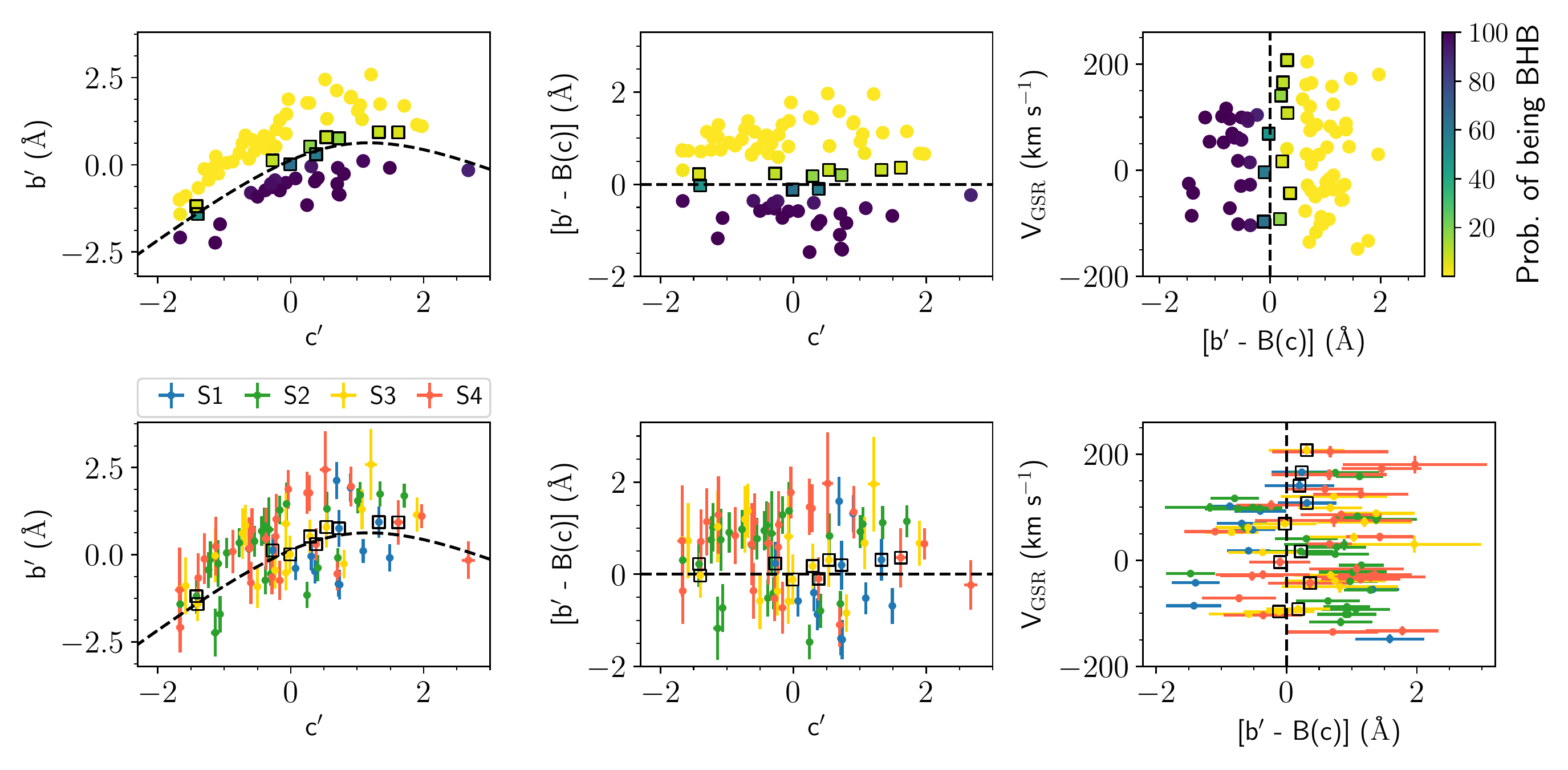}
  \caption{Stellar classification among BHB and BS stars from our
    FORS2 observations. The top row shows the target stars color-coded
    by to the probability of being classified as BHBs. At the bottom
    row, the stars are colour-coded according to the substructure they
    tentatively belong to: blue for S1, green for S2, yellow for S3
    and red for S4. Left panels show the $b^{\prime}$ and $c^{\prime}$
    coefficients from the S\'ersic profile, and the boundary derived
    based on SDSS data. In the middle panels, the $c^{\prime}$
    coefficient and the difference between the $b^{\prime}$ parameter
    and the corresponding $b^{\prime}$ from equation~\ref{eq:border}
    is shown. The rightmost panels present [$b^{\prime}$ - B(c)]
    against $V_{\rm GSR}$. Stars marked with open squares are those
    which can be either BHB or BS as they are located just at the
    separation boundary between the two classes.}
    \label{fig:cbparameters}
\end{figure*}

For our sample, the classification was carried out based on the
S\'ersic parameters for the same three Balmer lines available in the
SDSS spectra. The results of the fitting are shown in
Figure~\ref{fig:cbparameters}. Top row panels show the targets
colour-coded by the probability to belong to the BHB class according
to the division boundary given by the SVM. The dashed line shows the
boundary as defined in Eq.~\ref{eq:border}. Stars that are at the edge
have probabilities between the two classes and are marked with open
squares. They are listed as ``BHB/BS'' in Table~\ref{tab:targets}. In
the top left panel, the same parameter space as for SDSS stars is
shown. The separation shows that most of the stars can be classified
as either a BHB or a BS with high confidence. The top middle panel
shows the $c^{\prime}$ parameter as a function of the difference
between $b^{\prime}$ and Eq.~\ref{eq:border}. In this plane, BS stars
are clearly confined to large [$b^{\prime}$ - B(c)] values as compared
to the BHBs. The top right panel shows [$b^{\prime}$ - B(c)] versus
the Galactic Standard of Rest (GSR) velocities (derived from the
fitting of the six Balmer lines in the spectra). At the boundary,
there is a group of four stars with $V_{\rm GSR}$ greater than 100 km
s$^{-1}$ that cannot be unequivocally classified as BHB or BSs (marked
with open squares). The bottom row of panels in
Figure~\ref{fig:cbparameters} shows the distribution of $b^{\prime}$,
$c^{\prime}$ and $V_{\rm GSR}$ for our stars, colour-coded according
to the (tentative) substructure they belong to: blue for S1, green for
S2, yellow for S3 and red for S4. Stars from S4 tend to have the
largest error bars, likely because their spectra have lower S/N
(Sect.~\ref{sec:rvs}). At the boundary, three stars from S1, one from
S2, four from S3 and two stars from S4 can be either BHB, BS and,
therefore, are not excluded from the remaining analysis but marked in
every plot with an open square symbol.

\subsection{Heliocentric distances}

Once the stars have been classified as BHB or BSs, the heliocentric
distance can be derived accordingly. The distances for the BHB stars
were derived using the absolute magnitude equation given in
\cite{Be2016}. The dispersion of this relation is $\simeq$0.1 mag.

Absolute magnitudes for the BS candidates were determined considering
the $V$-band relation given in Eq. (14) by \cite{Kinman94}. To
transform from the Johnson $V$ magnitude into Sloan $g$, the colour
and magnitude relations from \cite{Jester2005} were used. Combining
all the relations, we derive the equation for the $g$-band absolute
magnitude of BSs as
\begin{equation}
 M_g = 2.2 + 4.557(g-r)_0 - 0.45{\rm [Fe/H]}.
\end{equation}\noindent
This relation is slightly different from the one derived by
\cite{Deason2011}. At a fixed metallicity, and depending on the
$(g-r)_0$ colour, the difference between the relations can be as high as
0.2 mag. Nonetheless, the dispersion of the relation from
\cite{Deason2011} is greater, of $\simeq$0.5 mag. To be conservative,
the same intrinsic dispersion is adopted for our relation, despite
taking into consideration the metallicity dependence on the absolute
magnitude of BSs. The distances for the BS candidates were derived
adopting a metallicity of [Fe/H]=$-1.5$. This value is a
compromise between the metallicities of the MW stellar halo, the LMC
and the SMC.

\begin{figure}
  \includegraphics[width=0.45\textwidth]{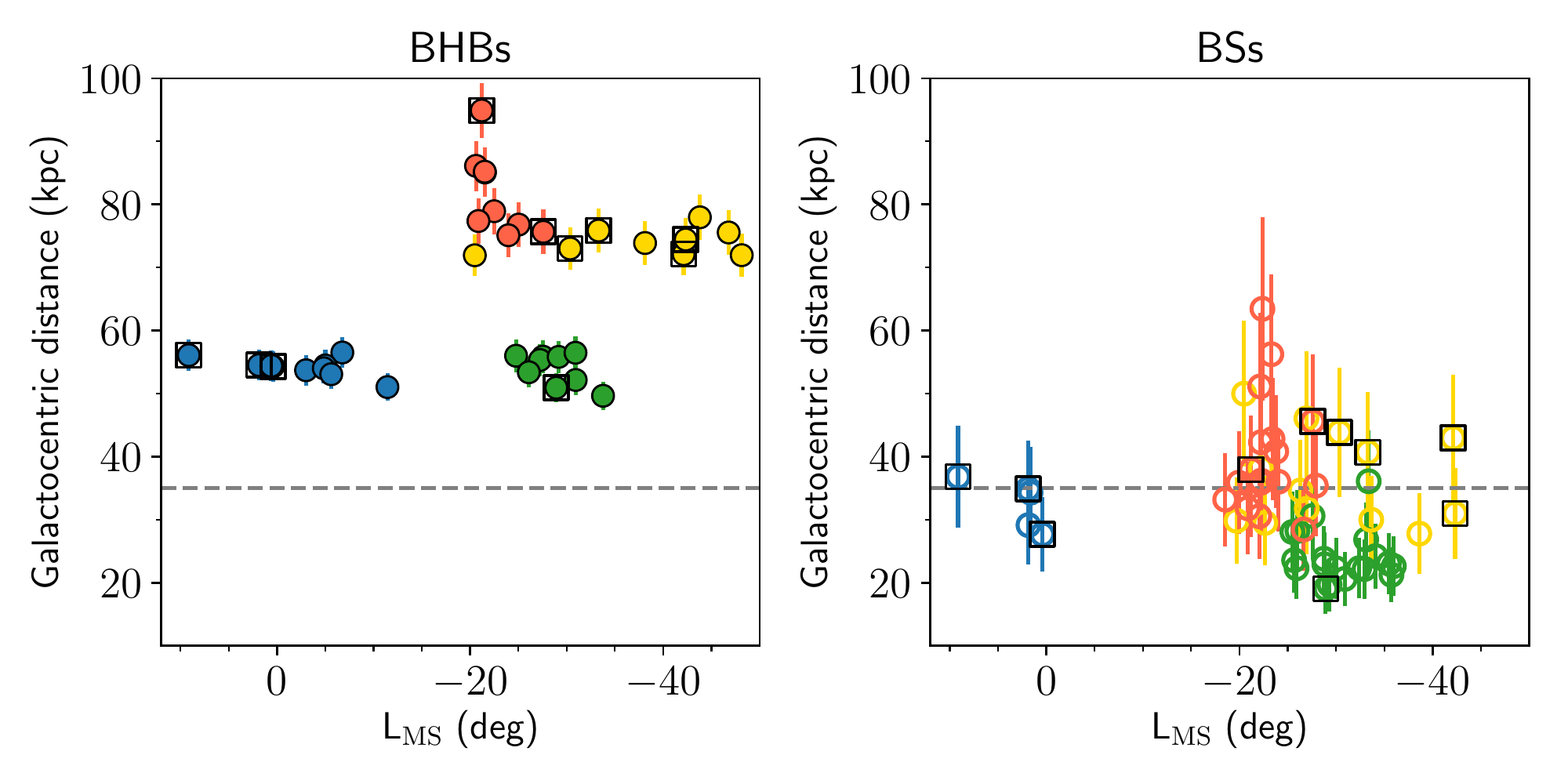}
  \caption{Galactocentric distances for BHB (left) and BS (right)
    stars as a function of $L_{\rm MS}$. The colours are the same as
    in Figure~\ref{fig:cbparameters}. Stars marked with open squares
    correspond to those that can be either BHB or BSs and therefore
    the corresponding distance in each case is considered. The dashed
    line marks the break radius at $\sim 35$ kpc. Stars with
    Galactocentric distances greater than this cut most likely belong
    to the LMC than to the MW halo.}
    \label{fig:distances}
\end{figure}

Figure~\ref{fig:distances} shows the Galactocentric distances of the
BHB and BS stars for the four different sub-structures as a function
of the MS longitude, $L_{\rm MS}$ \citep[defined by][]{Nidever08}. The
stars with uncertain classification were included in both panels, with
the according distance if they are considered BHB or BSs,
respectively, and marked with open squares. All of the BHBs are
located at $R > 40$~kpc, reaching out to $\simeq$100 kpc, far away
from the effective radius of the MW halo. For the BSs, the distances
are smaller, and most of them can be considered members of the MW
halo. Indeed, the typical break radius of the Galactic halo, beyond
which the stellar density falls-off more rapidly, is $\sim$25 -- 30
kpc \citep[see e.g.,][and references therein]{Deason2011, Xue2015,
  BlandHawthorn2016}. As a conservative cut, all the BSs at $R <$~35
kpc hereinafter will be considered to be part of the MW halo, while
those with $R \geq 35$~kpc are most likely members of the MCs. There
are one BS from S2, three from S3 and eleven from S4 that passed this
cut, and are included in the remaining analysis along with all the
BHBs.

Table~\ref{tab:targets} summarizes the properties of all A-type stars
with spectra from FORS2. The ID, equatorial coordinates (J2000.0) and
the apparent $g$ magnitude from the DES data release 1 (DR1) are
listed in columns (1)-(4); column (5) lists their GSR velocities, in
km s$^{-1}$, while the heliocentric distances, in kpc, are shown in
column (6). The distances were derived based on the classification of
the stars as BHB or BS and the adopted errors are the result of the
propagation of the uncertainty in the relation used: 0.1 mag for BHBs,
and 0.5 mag for BSs. For stars with no certain classification, the two
possible distances (if being a BHB or a BS) are reported. Column (7)
shows the equivalent width (EW) of the \caii\ K line. There is no EW
measurement for those stars in which the line was too shallow (see
Section~\ref{sec:calcium}). The class (BHB or BS) and the tentative
origin, MCs (for all the BHBs and BSs with Galactocentric
distances greater than 35 kpc), or MW (BSs with distances lower than
35 kpc) are listed in Columns (8) and (9),
  respectively. The reported velocities were first converted
  into heliocentric velocities using the barycentric correction from
  the {\tt rvcorrect} task of IRAF\footnote{IRAF is distributed by the
    National Optical Astronomy Observatories, which are operated by
    the Association of Universities for Research in Astronomy, Inc.,
    under cooperative agreement with the National Science Foundation.}
  and after that transformed to the Galactic Standard of Rest. The
  reported velocity errors do not consider possible systematic offsets
  due to imperfectly-centered position of the stars in the slit, which
  we estimate to be not larger than $\sim$ 15 km s$^{-1}$.

\subsection{Calcium as a proxy of metallicity}\label{sec:calcium}

Given the medium resolution of our spectra and the wavelength coverage
containing little other features apart from the Balmer lines, we do
not attempt to derive the stellar parameters of our target
stars. Nonetheless, most of the target spectra show the \caii\ K line
at $3933\,{\AA}$, particularly the BS stars. The EW of \caii\ line can
be used as a reliable metal abundance indicator in A-type stars
\citep[see e.g.,][]{Wilhelm99, Clewley2002, K11}. Therefore, we make
use of this line as a proxy for metallicity, as well as an additional
statistic to test for the presence of different stellar populations
among our targets.

\begin{figure}
\centering
  \includegraphics[width=0.45\textwidth]{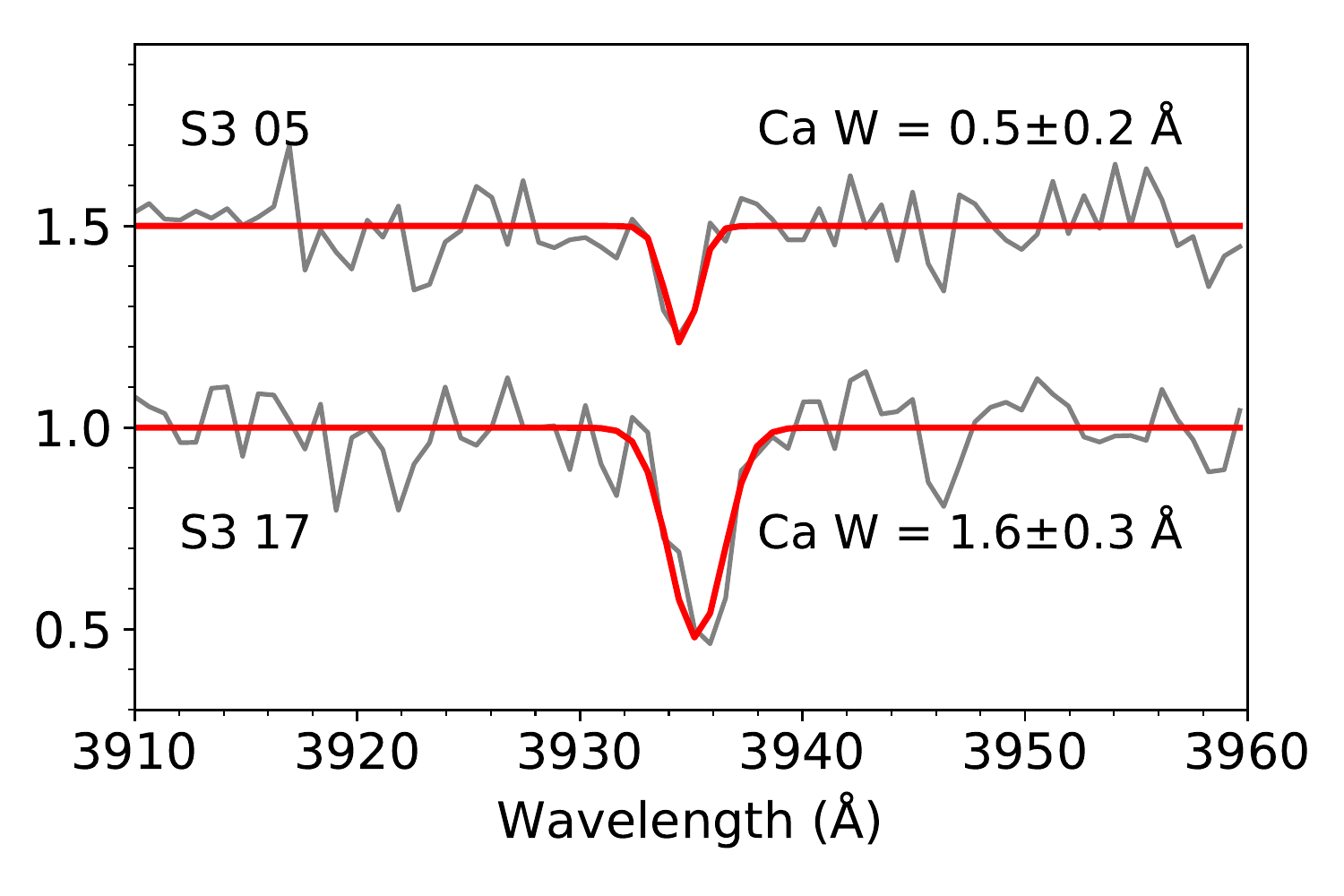}
  \caption{The normalized spectrum of S3~05 and S3~17 in the spectral range around the \caii\ K line are shown,
    vertically offset for clarity. The Gaussian fitting to the
    \caii\ K line is shown in the solid red lines. The EW of the
    \caii\ K line is at the right side of each spectrum.}
    \label{fig:ca_fit}
\end{figure}

To measure the EW of the \caii\ K line, a Gaussian fit to the
normalized spectrum was performed. From the width of the line the EW
was measured. Figure~\ref{fig:ca_fit} shows two examples, for S3 05
and S3 17, where the line is shallower (top) and deeper (bottom). The
spectra are vertically offset for clarity. The errors on the EW
measurements are the propagation of the errors on the fit parameters
for the line profile. Overall, the EWs tend to be higher for BSs
compared to BHB stars, in agreement with the fact that BHBs are
metal-poor stars.

\begin{figure}
  \includegraphics[width=0.48\textwidth]{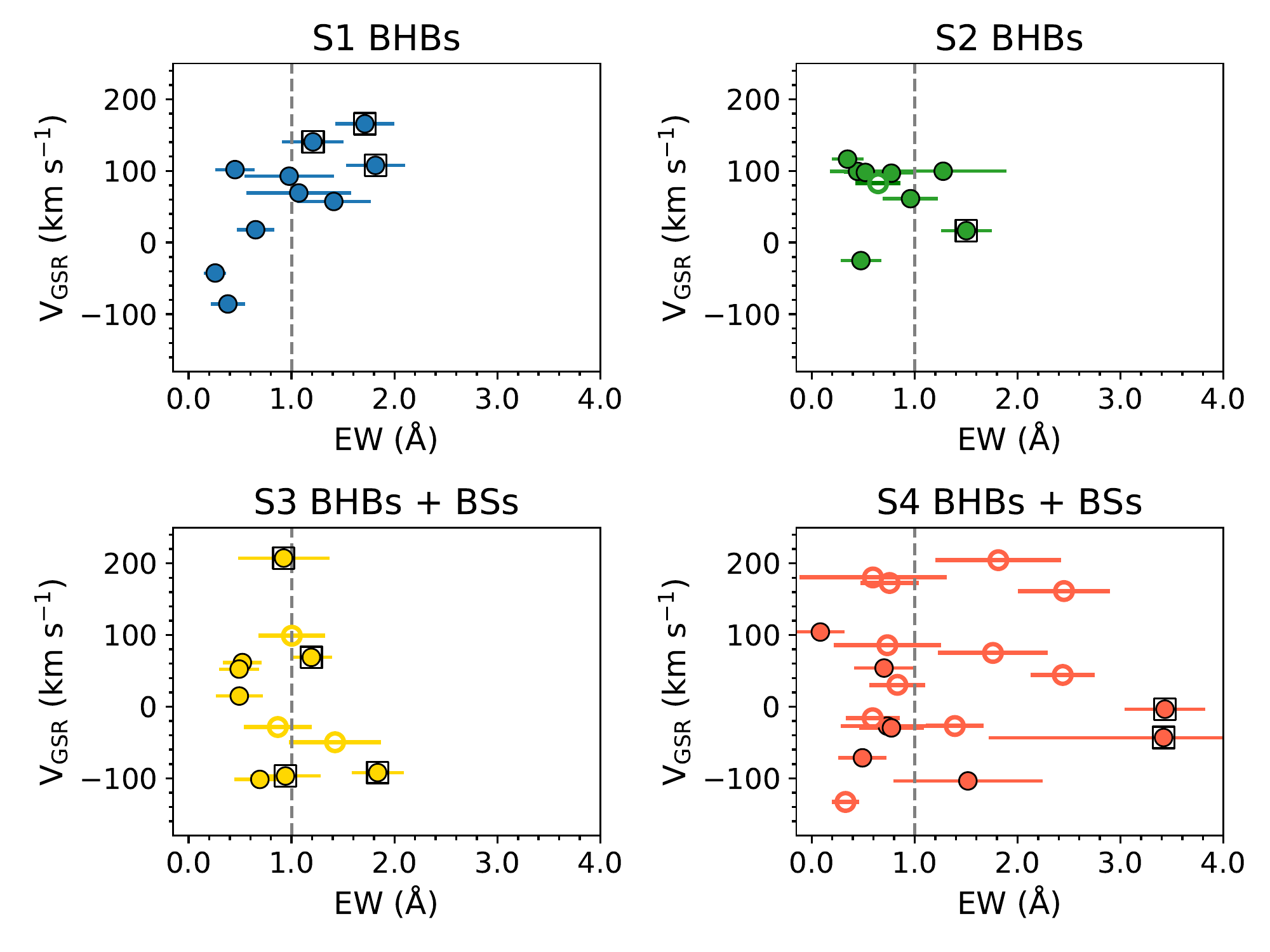}
  \caption{EW of the \caii\ K line as a function of the velocity
    $V_{\rm GSR}$ for the four different substructures. BHB and BS
    stars are marked as filled and open circles, respectively. The stars
    can be tagged as ``metal-rich'' and ``metal-poor'' at the boundary
    of EW = 1.0~{\AA} (vertical dashed line).}
    \label{fig:ews_vgsr}
\end{figure}

Figure~\ref{fig:ews_vgsr} shows the line-of-sight velocity as a
function of the \caii\ EW measurements. Only the stars classified as
BHBs (filled circles), the distant BSs (open circles) and those
without clear classification (open squares) are displayed (i.e., MW
stars are not included). The four panels correspond to the four
candidate substructures, as indicated on top of each panel. For S1
BHBs, there appears to be a correlation between the velocity and the
width of the \caii\ line or at least two groups with the EWs greater
or lower than 1.0 \AA. For S2, all BHBs have similar \caii\ EWs and,
moreover, all but one very similar V$_{\rm GSR}$ (with a
  velocity dispersion\footnote{Derived as the difference, in
    quadrature, of the standard deviation on the mean and the average
    error in velocity.} of $\sim$ 16 km s$^{-1}$). The fact that it
is not possible to claim any significant metallicity difference among
the S2 BHBs together with their relatively low velocity dispersion
supports the hypothesis of \citet{Be2016} that S2 is a stand-alone
substructure in the LMC halo. S3 and S4 stars (bottom row of panels)
are a mix of BHBs and BSs. In general, in BS stars, the EW of the
\caii\ line is higher, while the BHBs show a narrow range of the EW
measurements. This is not the case for one BHB/BS star in S3 with an
EW of $\sim$2.0 {\AA} and two BHB/BS from S4 with an EW$>$ 3.0 {\AA},
but these could be just because the stars are actually BSs despite the
classification was not certain. There is only one exception: one BHB
star in S4 which has EW$>$1.5 {\AA}, greater than any other BHB in
that substructure. Overall, we identify a difference in metallicity
only in S1 stars, which is reinforced by the two different velocity
trends analyzed in Section~\ref{sec:lmc_smc}.

To quantify the difference in metallicity between the ``metal-rich''
and the ``metal-poor'' BHB and BS stars, we derive empirical relations
between the Ca EW and [Fe/H] for a bigger sample of A-type stars,
following a similar procedure as the one used by
\cite{Wilhelm99}. Namely, A-type stars were recovered from SDSS DR9,
imposing the same colour cut in $(g-r)_0$ as the ones used to select
BHB candidates by \cite{Be2016}, and adding cuts in $(u-g)_0$ colour
and stellar parameters as follows:

\begin{eqnarray}
 0.9 < (u-g)_0 < 1.4 \text{,} \nonumber \\
 -0.30 < (g-r)_0 < -0.05 \text{,} \nonumber \\
 2.8 < \log{(g_s)} < 4.6 \text{,} \nonumber \\
 7500 \text{ K} < T_{\rm eff} < 9300 \text{ K,} \nonumber \\ 
 S/N > 30 \text{.}
\end{eqnarray}

\begin{figure*}
  \includegraphics[width=\textwidth]{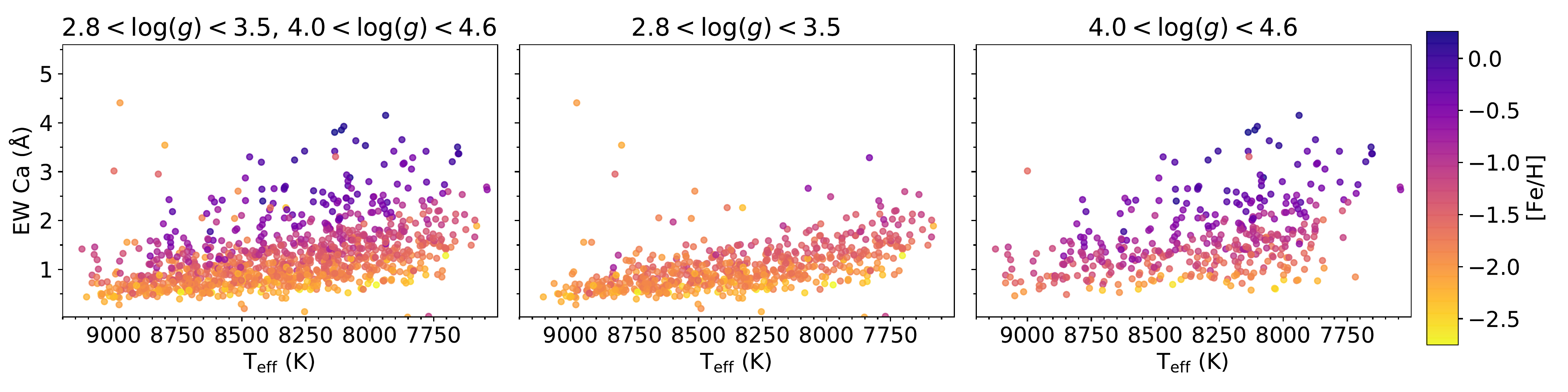}
  \caption{Equivalent widths of the \caii\ K line as a function of the
    effective temperature for A-type stars in SDSS DR9, colour-coded
    according to the metallicity [Fe/H]. From left to right, the
    logarithm of the surface gravity ranges are: 2.8 $< \log{(g_s)}
    <$3.5 and 4.0$< \log{(g_s)} <$4.6 (including both BHB and
      BS star candidates); 2.8$< \log{(g_s)} <$3.5 (BHB star
    candidates); 4.0 $< \log{(g_s)} <$ 4.6 (BS star candidates). For a
    given temperature, the scatter in the EW of Ca is due to the
    metallicity difference.}
    \label{fig:ca_teff_feh}
\end{figure*}

With this query, $\approx$1000 spectra were recovered. To be
consistent with the EW determination of the \caii\ line used above for
our target stars, the same approach was used for the SDSS A-type
stars, i.e. fitting a Gaussian to the Ca line profile in the
normalized spectrum. Using the spectroscopic stellar parameters from
the SDSS SEGUE Stellar Parameter Pipeline (see \citealt{Lee08} for
details), we investigated the correlation between the EW of Ca and
effective temperature, surface gravity and
[Fe/H]. Figure~\ref{fig:ca_teff_feh} shows the EW of Ca as a function
of the effective temperature for all the sample (left panel), stars
with $2.8 < \log{(g_s)} < 3.5$ (mainly BHB stars, middle panel), and
stars with higher surface gravities, $4.0 < \log{(g_s)} < 4.6$ (mainly
BSs, right panel), color-coded according to the [Fe/H] values. From
the figure, it is evident that the EW of Ca is a strong function of
effective temperature and [Fe/H], as previously claimed by
\cite{Wilhelm99} and \cite{Clewley2002}.  Moreover, the EW of Ca is
more sensitive to metallicity for cooler stars, whereas the EW of Ca
tends to zero for the hottest (bluest) stars. Given the marked
difference between BHBs and BSs, the relations converting EW of Ca
into [Fe/H] need to be treated separately.

\begin{figure*}
 \includegraphics[width=\textwidth]{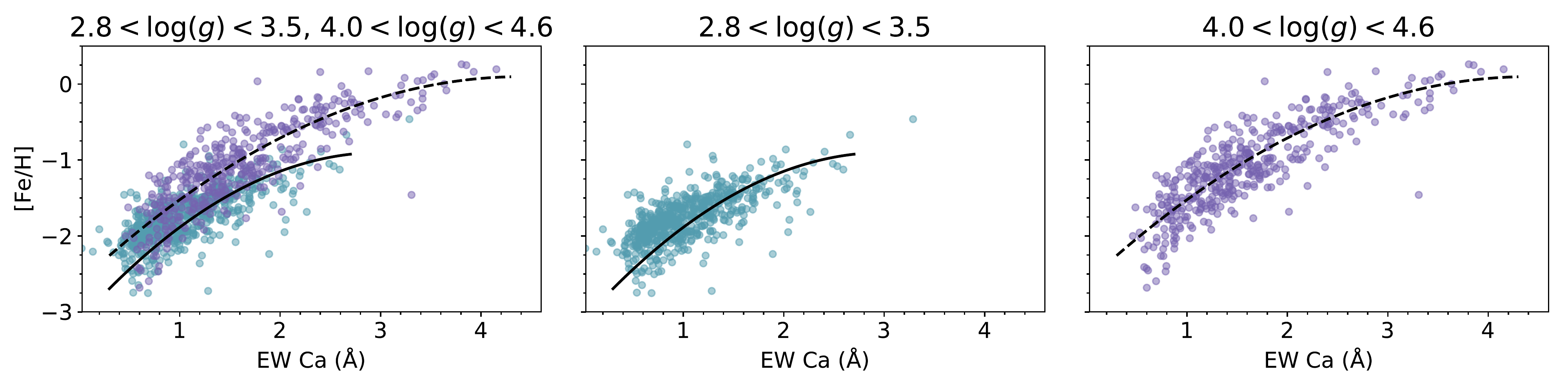}
 \includegraphics[width=\textwidth]{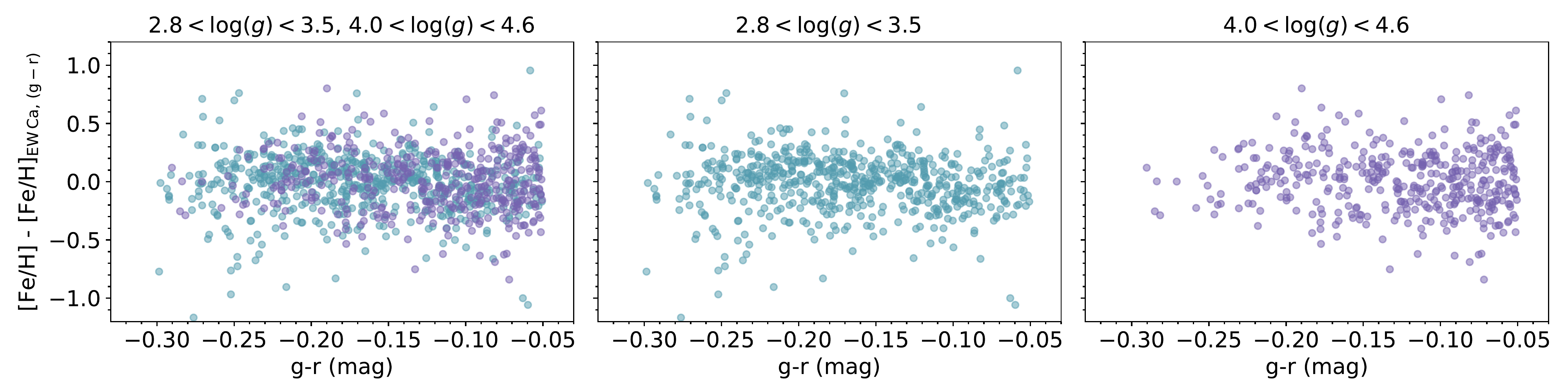}
 \caption{Top panels: Equivalent width of \caii\ K line versus
   metallicity [Fe/H] in the same three surface gravity ranges as in
   Figure~\ref{fig:ca_teff_feh}. The solid and dashed lines correspond
   to [Fe/H] as a function of the Ca EW relations derived for BHB
   and BS stars. Bottom panels show the residuals of the fit,
   including an additional colour term, as a function of the $(g-r)$
   colour, in the same three surface gravity ranges.}
 \label{fig:ca_feh_fitlines}
\end{figure*}

The quadratic relations fitted to the SDSS data are shown in the top
panels of Figure~\ref{fig:ca_feh_fitlines}, separately for the BHBs
and the BSs. For both stellar types, for a given value of EW of Ca,
there is a range of possible values of [Fe/H], indicative of the
dependence of the line shape properties on effective temperature. To
correct for this effect, we fit an additional $(g-r)_0$ color term to
the residuals. The relations to convert from Ca EW and
  $(g-r)_0$ into [Fe/H] for BHB and BS stars are the following

 \begin{eqnarray}
  {\rm [Fe/H]}_{\rm BHB} = -0.2\,\left(\text{EW Ca}\right)^2 + 1.47\,\left(\text{EW Ca}\right)  -1.77 \,(g-r) - 3.26 \label{feh_bhb}\text{,} \nonumber \\
  {\rm [Fe/H]}_{\rm BS} = -0.14 \,\left(\text{EW Ca}\right)^2 + 1.23 \,\left(\text{EW Ca}\right) - 2.08 \,(g-r) - 2.90 \label{feh_bs}\text{.}\nonumber
 \end{eqnarray}

The bottom panels in Figure~\ref{fig:ca_feh_fitlines} give the
residuals for the SDSS data and the relations derived, showing no
correlation with colour and a small dispersion around zero (1$\sigma$
dispersion of 0.25 dex for both BHB and BS
stars). Table~\ref{tab:s1_feh} shows the metallicity estimation for
our target stars based on the relations derived from the SDSS data
according to their classification as BHB or BS star. In the cases of
uncertain classification, metallicities using both BHB and BS
relations are presented. The errors were derived from the propagation
of the error on the EW of Ca measurement and the errors on the
coefficients of the relation used but do not consider the intrinsic
dispersion of the relations. For stars as hot as our targets,
  the SDSS pipeline relies heavily on the Ca II K line to derive
  metallicities given that only weak iron lines are available. Our
  approach is based on the same feature, the Ca II K line, which gives
  us confidence that our metallicity estimates are compatible with the
  SDSS [Fe/H] estimates.

Prior to deriving the metallicities from the Ca EWs and $(g-r)$, the
interstellar contribution to the EW of the Ca line was also derived
\citep{B90, K11}. The relations derived are based on the measured EW
of Ca and do not take into account the contribution from the
interstellar Ca along the line of sight.  However, we do not correct
for this contribution since the available relations
\citep[e.g.,][]{B90, K11} are derived based only on Galactic A-type
stars. As far as we are aware, there are no corrections for
interstellar Ca in stars belonging to the MCs. Using the relations for
Galactic stars given by \cite{K11}, we re-derive the relations
converting EWs into metallicities using SDSS stars, and our
metallicity estimates for our targets, finding that overall the
differences are small (below 0.1 dex). The small contribution of the
interstellar material on the measured Ca II line is also evident when
the derived [Fe/H] are compared to the extinction E(B-V) values for
our targets \citep[derived from the dust maps of][]{SF11}: there is no
correlation between both quantities, indicating that the measured EWs
in our targets are not highly affected by the interstellar material
along the line of sight. However, other effects could affect our
measurement of [Fe/H] using the Ca line such as accretion of
interstellar material or atomic diffusion \citep{Brown06}, in addition
to possible star-to-star variations in [$\alpha$/Fe]. Therefore, the
presented [Fe/H] values should be considered only as an approximate
estimate.  For the interested reader, a detailed study of
  interstellar Ca II absorption in the Milky Way was conducted by
  \cite{Murga15}.

\section{The LMC/SMC connection}\label{sec:lmc_smc}

\begin{figure}
 \includegraphics[width=0.48\textwidth]{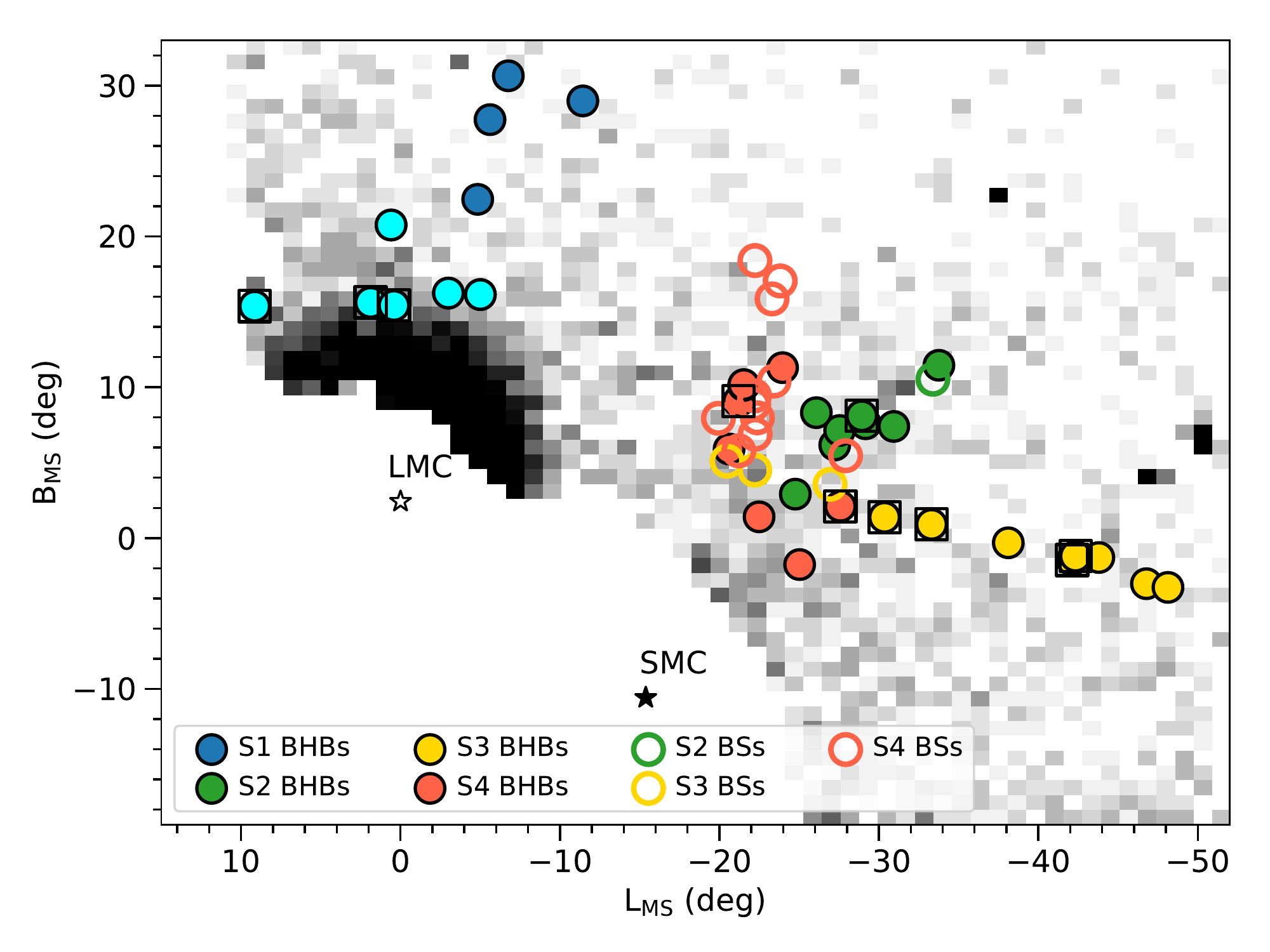}
 \caption{Spatial distribution in Magellanic coordinates (L${}_{\rm
     MS}$, B${}_{\rm MS}$) of BHB stars from the DES survey with our
   target stars overplotted with filled circles for BHB and open
   circles for BS stars. S1 stars with high and low relative
   metallicity are in cyan and blue, while S2, S3 and S4 stars
   are in green, yellow and red colours, respectively. The position of
   the LMC and SMC are marked with star symbols and labelled
   accordingly.}
 \label{fig:lms_bms}
\end{figure}

The celestial positions of our target stars, in Magellanic coordinates, are 
shown in Figure~\ref{fig:lms_bms}. Also shown in the figure, underlying the BHB 
and BS stars, is the density of the BHB candidate stars selected from the DES 
DR1 photometry using similar colour cuts as described in \cite{Be2016}: 
$(g-r, i-z)$ = (--0.40, --0.20), (--0.30, --0.15), (--0.25, --0.13), (--0.20, 
--0.12), (--0.13, --0.10), (--0.05, --0.09), (0.00, --0.09), (0.00, 0.00), 
(--0.10, 0.00), (--0.20, --0.02), (--0.30, --0.06), (--0.40, --0.14). From this 
density distribution alone, the more metal-rich BHBs from S1 (pale blue) can be 
associated with a disturbed portion of the LMC disk, while the rest of the group 
extends much further away, not only in Magellanic longitude but also in 
latitude, further away from the MS, while stars from S2, S3 and S4 are located 
closer to the MS but further away from the LMC itself.

\begin{table*}
\caption[]{A-type stars from the FORS2 observations.\label{tab:targets} }
\begin{tabular}{lcccccccc}
\hline \hline
  ID   & RA (J2000.0)  & DEC (J2000.0) & $g$    & $V_{\rm GSR}$ & $D$   & $W$ Ca & Class  & Group \\
       & (deg)         & (deg)         & (mag)  & (km s$^{-1}$) & (kpc) & (\AA)  &        &   \\
\hline
S1 01 & 69.47791       & --41.67166    & 19.12  &  --85.7 $\pm$  6.1 & 48  $\pm$ 2              &   0.4 $\pm$ 0.2  &  BHB     & MCs - M1  \\
S1 02 & 75.92958       & --44.00163    & 19.02  &    17.9 $\pm$  4.7 & 50  $\pm$ 2              &   0.7 $\pm$ 0.2  &  BHB     & MCs - M1  \\
S1 03 & 75.24083       & --40.96847    & 19.30  &  --42.5 $\pm$  5.2 & 53  $\pm$ 2              &   0.3 $\pm$ 0.1  &  BHB     & MCs - M1  \\
S1 16 & 98.09583       & --55.92491    & 19.39  &   140.6 $\pm$  8.5 & 55  $\pm$ 3/ 35 $\pm$ 8  &   1.2 $\pm$ 0.3  &  BHB/BS  & MCs - M1  \\
S1 28 & 85.51833       & --56.58208    & 19.30  &   107.7 $\pm$  7.8 & 53  $\pm$ 2/ 33 $\pm$ 8  &   1.8 $\pm$ 0.3  &  BHB/BS  & MCs - M1  \\
S1 30 & 85.39375       & --54.35069    & 19.07  &  --55.0 $\pm$  7.0 & 27  $\pm$ 6              &   0.5 $\pm$ 0.2  &  BS      & MW        \\
S1 32 & 82.94583       & --56.78797    & 19.12  &   165.8 $\pm$  6.8 & 53  $\pm$ 2/ 26 $\pm$ 6  &   1.7 $\pm$ 0.3  &  BHB/BS  & MCs - M1  \\
S1 38 & 77.15333       & --55.72430    & 19.13  &    57.4 $\pm$  5.4 & 52  $\pm$ 2              &   1.4 $\pm$ 0.4  &  BHB     & MCs - M1  \\
S1 43 & 73.76208       & --55.52858    & 19.23  &    69.3 $\pm$  6.4 & 53  $\pm$ 2              &   1.1 $\pm$ 0.5  &  BHB     & MCs - M1  \\
S1 50 & 83.41541       & --51.45541    & 19.11  &    92.7 $\pm$  5.6 & 52  $\pm$ 2              &   1.0 $\pm$ 0.4  &  BHB     & MCs - M1  \\
S1 57 & 84.92083       & --47.85800    & 19.22  & --148.2 $\pm$  8.3 & 31  $\pm$ 7              &   0.8 $\pm$ 0.3  &  BS      & MW        \\
S1 63 & 75.74666       & --49.33027    & 19.10  &   101.8 $\pm$  5.0 & 52  $\pm$ 2              &   0.5 $\pm$ 0.2  &  BHB     & MCs - M1  \\    
\hline
S2 01 & 30.75012       & --57.10719    & 19.49  &    99.4 $\pm$  8.2 & 57  $\pm$ 3              &   0.4 $\pm$ 0.3  &  BHB     & MCs - M2  \\
S2 02 & 32.34349       & --55.39111    & 19.30  & --116.7 $\pm$  7.9 & 28  $\pm$ 6              &   1.3 $\pm$ 0.3  &  BS      & MW        \\
S2 04 & 34.09958       & --53.90825    & 19.28  &   164.6 $\pm$  7.3 & 21  $\pm$ 5              &   2.2 $\pm$ 0.3  &  BS      & MW        \\
S2 05 & 35.28941       & --53.59127    & 19.20  &    29.6 $\pm$ 10.5 & 23  $\pm$ 5              &   1.4 $\pm$ 0.2  &  BS      & MW        \\
S2 08 & 32.28366       & --53.11711    & 19.19  &    99.6 $\pm$  5.5 & 55  $\pm$ 3              &   1.3 $\pm$ 0.6  &  BHB     & MCs - M2  \\
S2 09 & 33.14850       & --52.25447    & 19.25  &   116.5 $\pm$  5.5 & 56  $\pm$ 3              &   0.3 $\pm$ 0.2  &  BHB     & MCs - M2  \\
S2 10 & 30.53904       & --51.97269    & 19.22  &  --92.4 $\pm$  7.0 & 22  $\pm$ 5              &   1.1 $\pm$ 0.3  &  BS      & MW        \\
S2 12 & 31.98020       & --50.74277    & 19.15  &    96.9 $\pm$  6.3 & 56  $\pm$ 3              &   0.8 $\pm$ 0.2  &  BHB     & MCs - M2  \\
S2 13 & 31.93720       & --50.54833    & 18.99  &    11.6 $\pm$  6.7 & 18  $\pm$ 4              &   2.5 $\pm$ 0.3  &  BS      & MW        \\
S2 14 & 30.02499       & --49.50663    & 19.50  &  --25.2 $\pm$  5.2 & 52  $\pm$ 2              &   0.5 $\pm$ 0.2  &  BHB     & MCs - M1  \\
S2 15 & 33.94395       & --50.19563    & 19.34  &  --22.4 $\pm$  5.5 & 23  $\pm$ 5              &   0.7 $\pm$ 0.2  &  BS      & MW        \\
S2 17 & 31.46258       & --46.86097    & 19.07  &    76.7 $\pm$  8.7 & 21  $\pm$ 5              &   0.2 $\pm$ 0.5  &  BS      & MW        \\
S2 18 & 30.23741       & --46.61547    & 19.19  &   157.9 $\pm$  6.3 & 26  $\pm$ 6              &   0.9 $\pm$ 0.3  &  BS      & MW        \\
S2 19 & 31.68004       & --46.01044    & 19.10  &    24.9 $\pm$  6.7 & 21  $\pm$ 5              &   2.2 $\pm$ 0.3  &  BS      & MW        \\
S2 20 & 31.35933       & --45.65191    & 19.42  &    82.8 $\pm$  6.4 & 35  $\pm$ 8              &   0.6 $\pm$ 0.2  &  BS      & MCs - M2  \\
S2 21 & 32.33658       & --44.16247    & 19.14  &  --76.8 $\pm$  6.6 & 22  $\pm$ 5              &   0.6 $\pm$ 0.2  &  BS      & MW        \\
S2 25 & 32.24987       & --42.47661    & 19.30  & --101.8 $\pm$  5.6 & 21  $\pm$ 5              &   0.6 $\pm$ 0.2  &  BS      & MW        \\
S2 27 & 33.65883       & --40.94358    & 19.07  &  --39.7 $\pm$  6.1 & 20  $\pm$ 5              &   1.7 $\pm$ 0.3  &  BS      & MW        \\
S2 28 & 30.14699       & --55.76855    & 19.07  &  --87.7 $\pm$  6.4 & 28  $\pm$ 6              &   0.7 $\pm$ 0.2  &  BS      & MW        \\
S2 29 & 36.17920       & --52.50597    & 19.19  &    61.3 $\pm$  4.4 & 53  $\pm$ 2              &   1.0 $\pm$ 0.3  &  BHB     & MCs - M2  \\
S2 30 & 33.87925       & --51.79997    & 19.17  &   --9.4 $\pm$  6.4 & 30  $\pm$ 7              &   0.2 $\pm$ 0.2  &  BS      & MW        \\
S2 31 & 30.08287       & --50.65188    & 19.02  &  --56.1 $\pm$  6.3 & 21  $\pm$ 5              &   1.1 $\pm$ 0.2  &  BS      & MW        \\
S2 32 & 32.85958       & --50.57858    & 18.93  &    16.6 $\pm$  5.2 & 51  $\pm$ 2/ 18 $\pm$ 4  &   1.5 $\pm$ 0.2  &  BHB/BS  & MCs - M2  \\
S2 33 & 34.90295       & --46.74211    & 19.14  &    40.4 $\pm$  6.3 & 19  $\pm$ 4              &   2.3 $\pm$ 0.3  &  BS      & MW        \\
S2 34 & 32.03433       & --44.77616    & 18.92  &    98.0 $\pm$  4.8 & 49  $\pm$ 2              &   0.5 $\pm$ 0.2  &  BHB     & MCs - M2  \\
S2 35 & 33.82012       & --41.15047    & 19.00  & --101.8 $\pm$  6.3 & 19  $\pm$ 4              &   1.1 $\pm$ 0.2  &  BS      & MW        \\
\hline
S3 01 & 10.40004       & --44.47533    & 19.95  &  --92.0 $\pm$  7.3 & 73  $\pm$ 3/ 44 $\pm$ 10 &   1.8 $\pm$ 0.3 &  BHB/BS  & MCs - M1  \\
S3 02 & 9.692750       & --42.87016    & 19.98  &    14.9 $\pm$  7.2 & 79  $\pm$ 4              &   0.5 $\pm$ 0.2 &  BHB     & MCs - M2  \\
S3 03 & 10.61537       & --44.17408    & 19.84  &    68.8 $\pm$  5.3 & 76  $\pm$ 3/ 31 $\pm$ 7  &   1.2 $\pm$ 0.2 &  BHB/BS  & MCs - M2  \\
S3 05 & 6.025458       & --40.76222    & 19.99  &    61.6 $\pm$  6.6 & 77  $\pm$ 4              &   0.5 $\pm$ 0.2 &  BHB     & MCs - M2  \\
S3 06 & 5.101791       & --39.58691    & 19.90  &    52.5 $\pm$  6.5 & 73  $\pm$ 3              &   0.5 $\pm$ 0.2 &  BHB     & MCs - M2  \\
S3 08 & 39.95100       & --59.46427    & 19.89  &  --38.6 $\pm$  7.7 & 30  $\pm$ 7              &   0.8 $\pm$ 0.3 &  BS      & MW        \\
S3 09 & 39.27050       & --58.68730    & 20.07  &  --28.4 $\pm$  8.7 & 50  $\pm$ 12             &   0.9 $\pm$ 0.3 &  BS      & MCs - M1  \\
S3 10 & 36.12066       & --57.91583    & 19.97  &  --49.6 $\pm$ 11.9 & 38  $\pm$ 9              &   1.4 $\pm$ 0.4 &  BS      & MCs - M1  \\
S3 12 & 35.49362       & --57.60886    & 19.80  &    72.1 $\pm$  9.2 & 29  $\pm$ 7              &   1.2 $\pm$ 0.3 &  BS      & MW        \\
S3 15 & 29.44429       & --55.67291    & 19.92  &    30.1 $\pm$ 16.0 & 34  $\pm$ 8              &   ...           &  BS      & MW        \\
S3 16 & 29.18641       & --55.00816    & 20.01  &    99.0 $\pm$  8.2 & 46  $\pm$ 11             &   1.0 $\pm$ 0.3 &  BS      & MCs - M2  \\
S3 17 & 29.15583       & --54.97430    & 19.99  &    43.4 $\pm$  8.6 & 32  $\pm$ 7              &   1.6 $\pm$ 0.3 &  BS      & MW        \\
S3 19 & 22.72979       & --53.51613    & 19.97  &   207.4 $\pm$  8.5 & 74  $\pm$ 3/ 44 $\pm$ 10 &   0.9 $\pm$ 0.4 &  BHB/BS  & MCs - M2  \\
S3 22 & 14.34962       & --47.61008    & 19.83  & --101.5 $\pm$  7.3 & 75  $\pm$ 3              &   0.7 $\pm$ 0.2 &  BHB     & MCs - M1  \\
S3 23 & 14.72145       & --46.90730    & 19.81  &   120.1 $\pm$  7.4 & 28  $\pm$ 6              &   1.5 $\pm$ 0.4 &  BS      & MW        \\
S3 24 & 19.52979       & --51.26036    & 19.97  &  --96.7 $\pm$  7.1 & 77  $\pm$ 4/ 41 $\pm$ 9  &   0.9 $\pm$ 0.3 &  BHB/BS  & MCs - M1  \\
S3 25 & 19.34391       & --50.92566    & 19.90  &    88.3 $\pm$  7.8 & 30  $\pm$ 7              &   1.2 $\pm$ 0.3 &  BS      & MW        \\

\hline
S4 02 & 40.13054       & --58.00400    & 20.69  &  104.2 $\pm$  8.9 & 86  $\pm$ 4           &   0.1 $\pm$ 0.2 &  BHB     & MCs - M2  \\
S4 04 & 39.12958       & --57.69111    & 20.12  &   85.8 $\pm$  7.8 & 35  $\pm$ 8           &   0.7 $\pm$ 0.5 &  BS      & MCs - M2  \\
S4 06 & 43.52183       & --56.88783    & 20.28  &  204.5 $\pm$ 11.0 & 35  $\pm$ 8           &   1.8 $\pm$ 0.6 &  BS      & MCs - M2  \\
S4 12 & 39.12945       & --56.18783    & 20.19  &  161.4 $\pm$ 10.3 & 36  $\pm$ 8           &   2.5 $\pm$ 0.4 &  BS      & MCs - M2  \\
S4 16 & 40.25579       & --55.29122    & 20.60  &   30.5 $\pm$  6.2 & 63  $\pm$ 15          &   0.8 $\pm$ 0.3  & BS      & MCs - M2  \\
 \hline                                                                                                                                            
\end{tabular}                                                                                                                                                    
\end{table*}                                                                    %
\begin{table*}
\contcaption{A-type stars from the FORS2 observations. }
\begin{tabular}{lcccccccc}
\hline \hline
  ID   & RA (J2000.0)  & DEC (J2000.0) & $g$   &  $V_{\rm GSR}$ & $D$   & $W$ Ca & Class  & Group \\
       & (deg)         & (deg)         & (mag) & (km s$^{-1}$) & (kpc) & (\AA)  &        &   \\
\hline
S4 17 & 43.06666       & --55.22061    & 20.32  &  --3.5 $\pm$  7.9 & 95  $\pm$ 4/ 37 $\pm$ 9  &   3.4 $\pm$ 0.4  &  BHB/BS  & MCs - M1  \\
S4 18 & 41.58354       & --54.76944    & 20.35  &  180.4 $\pm$ 17.0 & 51  $\pm$ 12             &   0.6 $\pm$ 0.7  &  BS      & MCs - M2  \\
S4 22 & 45.46570       & --48.95708    & 20.20  & --31.0 $\pm$  9.0 & 34  $\pm$ 8              &   2.1 $\pm$ 0.4  &  BS      & MW        \\
S4 26 & 47.22491       & --48.60963    & 20.70  &   44.3 $\pm$  8.1 & 55  $\pm$ 13             &   2.4 $\pm$ 0.3  &  BS      & MCs - M2  \\
S4 28 & 47.69408       & --47.35980    & 20.62  &   75.0 $\pm$ 11.9 & 39  $\pm$ 9              &   1.8 $\pm$ 0.5  &  BS      & MCs - M2  \\
S4 29 & 50.65604       & --47.11874    & 20.54  &  172.6 $\pm$  8.8 & 40  $\pm$ 9              &   0.8 $\pm$ 0.3  &  BS      & MCs - M2  \\
S4 31 & 31.35824       & --59.81124    & 20.15  &   53.9 $\pm$  6.9 & 80  $\pm$ 4              &   0.7 $\pm$ 0.3  &  BHB     & MCs - M2  \\
S4 32 & 36.47429       & --59.64105    & 19.91  &  133.9 $\pm$  8.5 & 32  $\pm$ 7              &   2.0 $\pm$ 0.4  &  BS      & MW        \\
S4 38 & 23.29837       & --59.68866    & 19.97  &--103.6 $\pm$  7.4 & 78  $\pm$ 4              &   1.5 $\pm$ 0.7  &  BHB     & MCs - M1  \\
S4 46 & 27.63279       & --56.05813    & 19.83  &--135.1 $\pm$  4.9 & 28  $\pm$ 7              &   1.3 $\pm$ 0.3  &  BS      & MW        \\
S4 47 & 26.47745       & --55.38130    & 20.05  & --43.1 $\pm$ 10.9 & 76  $\pm$ 4/ 46 $\pm$ 11 &   3.4 $\pm$ 1.7  &  BHB/BS  & MCs - M1  \\
S4 61 & 30.65258       & --53.06022    & 20.02  & --16.1 $\pm$ 10.1 & 35  $\pm$ 8              &   0.6 $\pm$ 0.3  &  BS      & MCs - M1  \\
S4 64 & 49.46729       & --54.68383    & 19.83  & --36.0 $\pm$  8.1 & 32  $\pm$ 7              &   0.6 $\pm$ 0.2  &  BS      & MW        \\
S4 67 & 43.79241       & --54.21013    & 20.08  & --27.1 $\pm$  7.6 & 85  $\pm$ 4              &   0.7 $\pm$ 0.5  &  BHB     & MCs - M1  \\
S4 68 & 42.14212       & --54.31499    & 19.92  &--133.0 $\pm$  8.2 & 35  $\pm$ 8              &   0.3 $\pm$ 0.1  &  BS      & MCs - M1  \\
S4 72 & 43.83629       & --54.15716    & 20.17  & --29.6 $\pm$  8.6 & 85  $\pm$ 4              &   0.8 $\pm$ 0.3  &  BHB     & MCs - M1  \\
S4 76 & 43.24604       & --53.76177    & 19.99  &  124.5 $\pm$  7.9 & 29  $\pm$ 7              &   0.5 $\pm$ 0.4  &  BS      & MW        \\
S4 77 & 41.68458       & --52.77966    & 20.06  & --27.0 $\pm$  9.0 & 42  $\pm$ 10             &   1.4 $\pm$ 0.3  &  BS      & MCs - M1  \\
S4 78 & 41.99187       & --51.76663    & 19.85  & --71.4 $\pm$  7.4 & 75  $\pm$ 3              &   0.5 $\pm$ 0.2  &  BHB     & MCs - M1  \\
 \hline                                                                                                                                            
\end{tabular}                                                                                                                                                    
\end{table*}

Given the clumpy distribution of the Magellanic BHB and BS stars in the 
phase-space, we seek to compare our measurements to the expectations for the LMC 
halo and the previous detections. In particular, \cite{Munoz2006} identified a 
group of 15 giant stars in the area near the Carina dwarf spheroidal satellite 
galaxy, with a mean heliocentric velocity of 332 km s$^{-1}$. This moving group 
of stars also have metallicities, colours and magnitudes consistent with the red 
clump of the LMC. These stars were found at $\sim20$ degrees from the LMC centre, 
and their GSR velocities are in good agreement with the extrapolated velocity 
trend expected for LMC halo stars \citep[see Figure~17 in][]{Munoz2006}.

\begin{table}
 \centering
\caption[]{Metallicity estimates based on the EW of Ca and $(g-r)$ color.\label{tab:s1_feh} }
\begin{tabular}{lccc}
\hline \hline
  ID   & $W$ Ca        & Class  & [Fe/H]\\
       & (\AA)         &        &  \\
\hline                                        
S1 01 & 0.4 $\pm$ 0.2   &  BHB    &  --1.84 $\pm$ 0.41 \\
S1 02 & 0.7 $\pm$ 0.2   &  BHB    &  --1.73 $\pm$ 0.43 \\
S1 03 & 0.3 $\pm$ 0.1   &  BHB    &  --2.03 $\pm$ 0.37 \\

S1 16 & 1.2 $\pm$ 0.3   &  BHB/BS &  --0.99 $\pm$ 0.59 / --0.63 $\pm$ 0.42  \\
S1 28 & 1.8 $\pm$ 0.3   &  BHB/BS &  --0.56 $\pm$ 0.77 / --0.15 $\pm$ 0.47  \\
S1 32 & 1.7 $\pm$ 0.3   &  BHB/BS &  --0.81 $\pm$ 0.72 / --0.45 $\pm$ 0.45  \\
S1 38 & 1.4 $\pm$ 0.4   &  BHB    &  --0.95 $\pm$ 0.65  \\
S1 43 & 1.1 $\pm$ 0.5   &  BHB    &  --1.17 $\pm$ 0.69  \\
S1 50 & 1.0 $\pm$ 0.4   &  BHB    &  --1.39 $\pm$ 0.63  \\
S1 63 & 0.5 $\pm$ 0.2   &  BHB    &  --1.95 $\pm$ 0.42  \\    
 \hline                                         
S2 01 & 0.4 $\pm$ 0.3  &  BHB    &  --1.85 $\pm$ 0.49  \\
S2 08 & 1.3 $\pm$ 0.6  &  BHB    &  --1.20 $\pm$ 0.75  \\
S2 09 & 0.3 $\pm$ 0.2  &  BHB    &  --2.08 $\pm$ 0.39  \\
S2 12 & 0.8 $\pm$ 0.2  &  BHB    &  --2.00 $\pm$ 0.47  \\
S2 14 & 0.5 $\pm$ 0.2  &  BHB    &  --1.61 $\pm$ 0.45  \\
S2 20 & 0.6 $\pm$ 0.2  &  BS     &  --1.16 $\pm$ 0.35  \\
S2 29 & 1.0 $\pm$ 0.3  &  BHB    &  --1.32 $\pm$ 0.53   \\
S2 32 & 1.5 $\pm$ 0.2  &  BHB/BS &  --1.35 $\pm$ 0.64 / --1.08 $\pm$ 0.40  \\
S2 34 & 0.5 $\pm$ 0.2  &  BHB    &  --1.97 $\pm$ 0.43  \\
\hline                                          
S3 01 & 1.8 $\pm$ 0.3  &  BHB/BS & --0.57 $\pm$ 0.77  / --0.17 $\pm$ 0.46  \\
S3 02 & 0.5 $\pm$ 0.2  &  BHB    & --1.93 $\pm$ 0.45  \\
S3 03 & 1.2 $\pm$ 0.2  &  BHB/BS & --1.41 $\pm$ 0.54  / --1.12 $\pm$ 0.36  \\
S3 05 & 0.5 $\pm$ 0.2  &  BHB    & --1.79 $\pm$ 0.43  \\
S3 06 & 0.5 $\pm$ 0.2  &  BHB    & --1.79 $\pm$ 0.43  \\
S3 09 & 0.9 $\pm$ 0.3  &  BS     & --0.90 $\pm$ 0.43  \\
S3 10 & 1.4 $\pm$ 0.4  &  BS     & --0.75 $\pm$ 0.51  \\    
S3 16 & 1.0 $\pm$ 0.3  &  BS     & --0.81 $\pm$ 0.43  \\
S3 19 & 0.9 $\pm$ 0.4  &  BHB/BS & --1.26 $\pm$ 0.64  / --0.89 $\pm$ 0.52  \\
S3 22 & 0.7 $\pm$ 0.2  &  BHB    & --1.86 $\pm$ 0.47   \\
S3 24 & 0.9 $\pm$ 0.3  &  BHB/BS & --1.39 $\pm$ 0.57  / --1.05 $\pm$ 0.44  \\   
\hline                                         
S4 02 & 0.1 $\pm$ 0.2  &  BHB     & --2.01 $\pm$ 0.49  \\
S4 04 & 0.7 $\pm$ 0.5  &  BS      & --1.57 $\pm$ 0.59  \\
S4 06 & 1.8 $\pm$ 0.6  &  BS      & --0.77 $\pm$ 0.60  \\
S4 12 & 2.5 $\pm$ 0.4  &  BS      & --0.29 $\pm$ 0.60  \\
S4 16 & 0.8 $\pm$ 0.3  &  BS      & --0.90 $\pm$ 0.39  \\
S4 17 & 3.4 $\pm$ 0.4  &  BHB/BS  & --0.69 $\pm$ 1.61 /   0.15 $\pm$ 0.82  \\
S4 18 & 0.6 $\pm$ 0.7  &  BS      & --1.31 $\pm$ 0.80  \\
S4 26 & 2.4 $\pm$ 0.3  &  BS      &   0.05 $\pm$ 0.58  \\
S4 28 & 1.8 $\pm$ 0.5  &  BS      & --0.85 $\pm$ 0.56  \\
S4 29 & 0.8 $\pm$ 0.3  &  BS      & --1.72 $\pm$ 0.38  \\
S4 31 & 0.7 $\pm$ 0.3  &  BHB     & --1.56 $\pm$ 0.52   \\
S4 38 & 1.5 $\pm$ 0.7  &  BHB     & --0.96 $\pm$ 0.82   \\
S4 47 & 3.4 $\pm$ 1.7  &  BHB/BS  & --0.25 $\pm$ 1.65 /   0.66 $\pm$ 0.94  \\
S4 61 & 0.6 $\pm$ 0.3  &  BS      & --1.65 $\pm$ 0.37  \\
S4 67 & 0.7 $\pm$ 0.5  &  BHB     & --1.99 $\pm$ 0.63 \\
S4 68 & 0.3 $\pm$ 0.1  &  BS      & --1.85 $\pm$ 0.27  \\
S4 72 & 0.8 $\pm$ 0.3  &  BHB     & --1.58 $\pm$ 0.53 \\
S4 77 & 1.4 $\pm$ 0.3  &  BS      & --0.59 $\pm$ 0.42  \\
S4 78 & 0.5 $\pm$ 0.2  &  BHB     & --2.03 $\pm$ 0.45  \\                  
\hline                 
\end{tabular}                                                                                                                                                   
\end{table}   

\begin{figure}
  \includegraphics[width=0.48\textwidth]{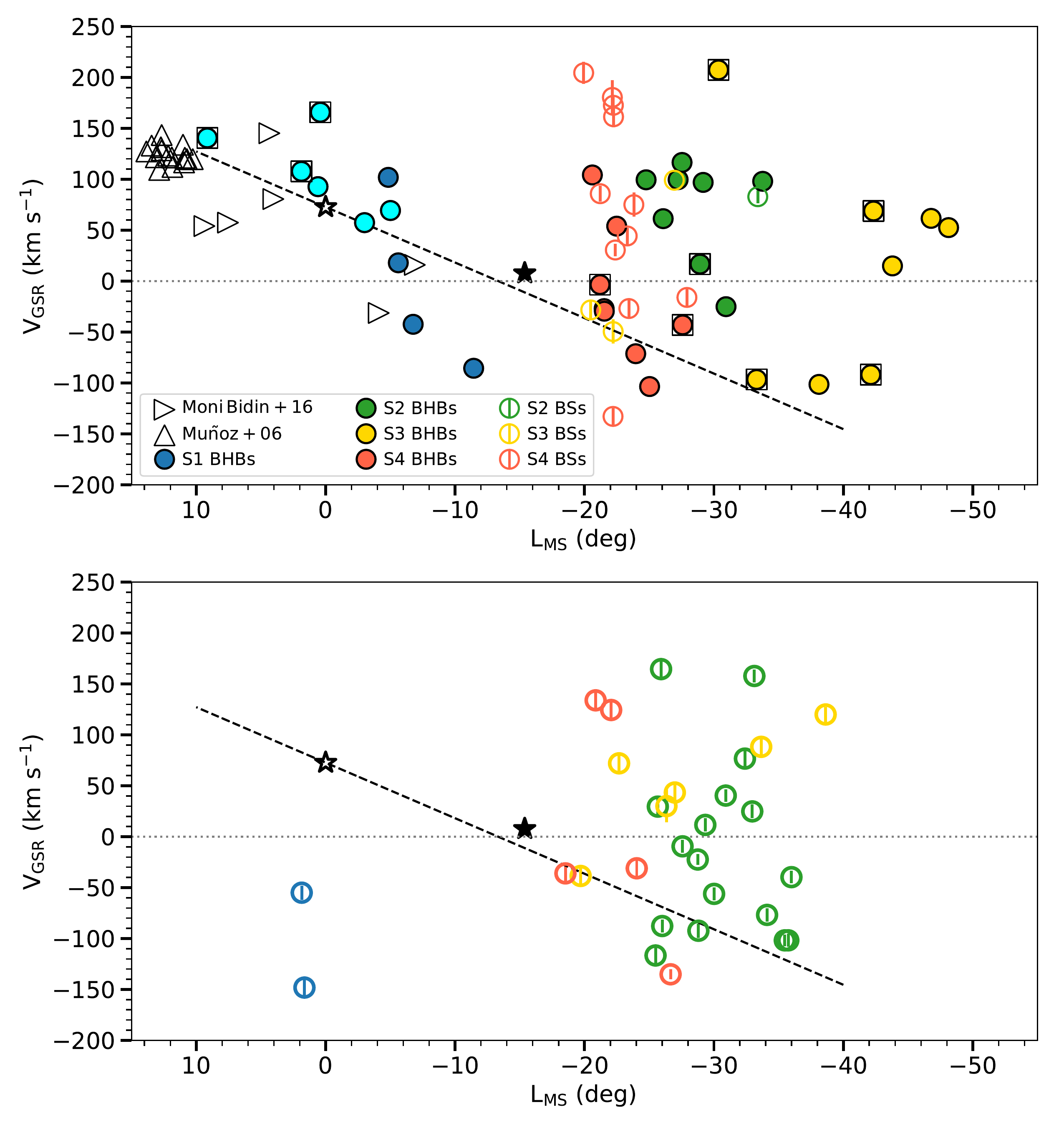}
  \caption{Phase-space distribution of the Magellanic BHB (filled
    circles) and BS (open circles) stars. The colours are the same as
    in Figure~\ref{fig:lms_bms}. The dashed line shows the projected
    LMC's velocity vector onto the line of sight crossing the LMC halo
    at $L_{\rm MS} = 0\degr$ while the dotted line marks the expected
    GSR velocity for MW stars. The open and filled stars mark the
    position of the LMC and SMC, respectively. The top panel shows the
    BHBs and distant BSs for the four streams, while the bottom panel
    shows those BS stars with smaller distances, and that most likely
    belong to the MW halo. Black open triangles correspond to giant
    stars in the 332 km s$^{-1}$ group identified by
    \citet{Munoz2006}, while the open rightwards-pointing triangles
    are young stars in the periphery of the LMC detected by
    \citet{MoniBidin17}.}
    \label{fig:lms_vgsr}
\end{figure}

The phase-space distributions of the Magellanic BHB and BS stars and
the giant stars detected by \cite{Munoz2006} are shown in the top
panel of Figure~\ref{fig:lms_vgsr}. The bottom panel includes only the
BS stars with Galactocentric distances $R \leq 35$ kpc. In the top
panel, S1 stars are colour-coded according to the EW of the
\caii\ line: those with EWs greater than 1.0~{\AA} are in cyan, while
the stars with smaller EWs are in blue. This separation is based on
the fact that those stars with EWs larger than 1.0~{\AA} have [Fe/H]
$\geq -1.4$, while the rest of the S1 stars have metallicities
[Fe/H]~$< -1.4$. The LMC's mean velocity vector projected onto the
line of sight crossing the LMC halo at $L_{\rm MS} = 0\degr$ is shown
as a dashed line. The relatively more metal-rich S1 BHBs nicely agree
with the measurements of \cite{Munoz2006} and the velocity gradient
expected for the LMC halo. In contrast, the metal-poor S1 BHBs follow
a different (much steeper) velocity gradient. The trend is not
dissimilar to the measurements at $L_{\rm MS} < 0\degr$ for young
stars in the periphery of the LMC recently measured by
\cite{MoniBidin17}. However, our stars are metal-poor and not
consistent with stars formed in-situ as in the case of
\cite{MoniBidin17}. We speculate that the metal-poor S1 BHB stars
could belong to a stream located inside the halo of the LMC, but
further observations of S1 stars are needed to confirm this
hypothesis. BHB and BS stars from S2, S3 and S4 with $V_{\rm GSR} < 0$
km s$^{-1}$ roughly follow the LMC halo velocity trend.  Their angular
separations from the LMC range from $\sim 20$ to 42 deg, much farther
away than any previous detections of LMC stars. In the bottom panel,
the phase-space distribution of the BS stars likely belonging to the
MW does not show any velocity trend and seems to be consistent with
the MW's halo velocity distribution, centered at $V_{\rm GSR}$ = 0 km
s$^{-1}$ and with $\sigma = 100$ km s$^{-1}$.

The six BHB stars and the one BS star from S2 with $V_{\rm GSR} > 0$
km s$^{-1}$ have a mean velocity of $V_{\rm GSR} = 93$ km s$^{-1}$,
and a velocity dispersion of $\sim$15 km s$^{-1}$. This dynamically
cold group is also very concentrated in Galactocentric distance, as
seen in Figure~\ref{fig:distances}, as well as appearing as a thin
stream on the sky (see Figure~\ref{fig:lms_bms}).  All stars in S2 are
metal-poor, with an average metallicity for the six BHB stars of
[Fe/H] = --1.7 $\pm$~0.3 dex, which reinforces the hypothesis that
they belong to a single substructure.

Among the S3 BHBs, S3~19 has the highest value of the line-of-sight
velocity, in fact it is the largest velocity overall in our
sample. This star lies on the decision boundary which separates the
BHB and the BS stars. Discarding this particular case, some of the S3
stars follow the LMC halo velocity gradient, reaching out to 42 deg from
the LMC. There is another group of S3 stars at $L_{\rm MS} = -40$ to
$-50$ deg, and $V_{\rm GSR} \sim 50$ km s$^{-1}$, that seems to be the
continuation of the S2 group. In other words, the velocities for BSs
in S3 are in good agreement with the velocities derived for BHBs in S2
and S4. Note however, that this similarity between S2 and S3 stars is
only apparent in this particular projection of the phase-space, and
not on the sky.

Stars from S4 (both BHBs and BSs) appear roughly following the LMC's
halo trend with $V_{\rm GSR} < -50$ km s$^{-1}$ at $L_{\rm MS}$ =
--20$\degr$ to --30$\degr$.  There are also S4 stars with radial
velocities $V_{\rm GSR} > 0$ km s$^{-1}$, reaching up to $V_{\rm GSR}$
= 205 km s$^{-1}$ (S4~06), reinforcing the idea that, given their
distances and velocities, these stars are possibly coming from the MCs
than from the MW. Nonetheless, we can not exclude the possibility that
some, plausibly a large fraction, of these stars can actually be part
of the MW halo. This is because the kinematic model predictions for
the virialised Galactic stellar halo and the Magellanic debris overlap
significantly in this region of the sky. At distances beyond 60 kpc,
the V$_{\rm GSR}$ distribution of the MW stellar halo can be described
by a Gaussian with a mean at zero and a dispersion of 90 $\pm$ 20 km
s$^{-1}$ \citep[see][]{Deason2012}. Therefore, in what follows we do
not claim an unambiguous detection of Magellanic debris for $L_{\rm
  MS} < -20^{\circ}$, but rather explore the trends under the
assumption that some of the stars in our sample could genuinely come
from the Clouds.

Interestingly, there is a group of stars spanning at least 10 deg on
the sky (with $L_{\rm BS}$ from --20$\degr$ to --35$\degr$) that show
very similar radial velocities, clumped at $V_{\rm GSR} \approx$ 84.0
km s$^{-1}$ (including stars from S2, S3 and S4 with $V_{\rm GSR}$
between 50 and 150 km $s^{-1}$).  This cold (velocity dispersion of
18.0 km s$^{-1}$) group does not seem to be related either to the
LMC's halo velocity trend at this position or to the MW halo. The
possible origin of this group is discussed in the following Section.

\begin{figure*}
 \includegraphics[width=\textwidth]{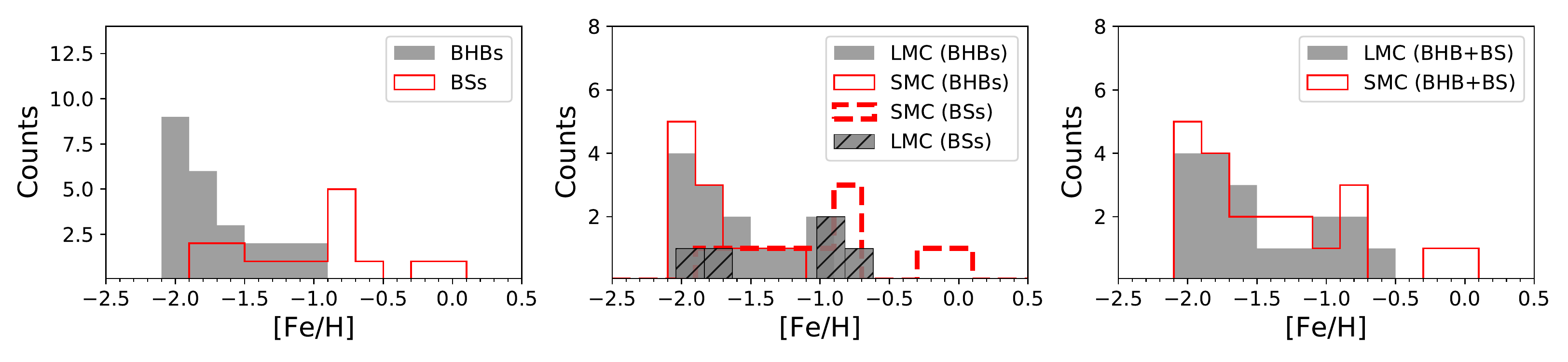}
 \caption{Metallicity distribution of BHB and BS stars in the
   MCs. Left: [Fe/H] distribution of BHB (grey filled bars) and BS
   (red unfilled bars) samples, irrespective of their possible
   association with the LMC or SMC. Middle: Same stars as in the left
   panel but divided into four subsamples: BHB and BS stars likely
   belonging to the LMC (grey filled and hatched bars, respectively),
   BHB (red unfilled bars) and BS (red dashed-line bars) stars likely
   associated to the SMC.  Right: Metallicity distribution for the
   same stars according to the association to the LMC (grey filled
   bars) or the SMC (red unfilled bars), irrespective of their stellar
   classification.}
 \label{fig:feh_histograms}
\end{figure*}

Figure~\ref{fig:feh_histograms} shows the metallicity distributions
for stars belonging to the four different streams that are likely
members of the MCs. Stars without EW of Ca measured (too shallow
lines) and for which only upper limits in [Fe/H] are estimated are not
included in the histograms. The left panel shows the distribution of
BHB and BS stars, irrespective of their membership; the middle panel
shows the same stars but divided into four subsamples: (i) BHB and
(ii) BS stars likely belonging to the LMC, (iii) BHB and (iv) BS stars
likely belonging to the SMC based on their measured radial velocities;
while the right panel shows the metallicity distribution of LMC and
SMC stars (irrespective of their classification as BHB or BS). The
LMC/SMC classification is tentative and is based on the comparison
with numerical simulations as explained in
Section~\ref{sec:simulations}. The metallicity distribution of the BHB
stars shows that the population is metal-poor, with an average [Fe/H]
= --1.69 and 1$\sigma$ dispersion of 0.34 dex. The high metallicity of
S4~17 and S4~47 can be explained considering they are more likely a BS
or an A-type star instead of being a BHB. In fact, \cite{Clewley2002}
found that a few A-type or BS stars can be misclassified as BHB stars
using the scale-width-shape method if those stars have anomalously
high metallicity ([Fe/H] $>$ --0.5). Since S4~17 and S4~47 have very
uncertain [Fe/H] values (given their high EW of Ca), no firm
conclusions about their classification can be established. For the BS
stars, the metallicity distribution tend to be more metal-rich, with a
mean metallicity of [Fe/H] = --0.75 and 1$\sigma$ dispersion of 0.36
dex (excluding the four most metal-poor stars, with [Fe/H] < --1.5
dex). The two most metal-rich stars deviates from the group, with
[Fe/H] = --0.3 (S4~12) and 0.05 dex (S4~26). It could be possible that
this group of stars corresponds to bona-fide (young) main sequence
stars, instead of BSs, which occupy the same region in the
colour-magnitude diagram and should have similar surface gravity as
dwarf stars. The same explanation can be true for the stars
classified as BHB/BS (not included in the metallicity distributions
shown in Figure~\ref{fig:feh_histograms}) which have [Fe/H] up to 0.66 
dex (S4~47).

%


The middle panel in Figure~\ref{fig:feh_histograms} shows the
metallicity distribution or the BHB and BS stars that are likely
associated with the LMC or SMC separately. Most of the LMC stars
(18/5) are BHBs with a mean metallicity [Fe/H] = --1.62 and a
1$\sigma$ dispersion of 0.38 dex, while the BHB and BS stars
associated to the SMC have slightly lower mean metallicity, [Fe/H] =
--1.77 for BHBs (1$\sigma$ = 0.28 dex) and [Fe/H] = --0.93 for BSs
(1$\sigma$ = 0.51 dex). As a whole sample, the mean metallicity of SMC
stars is [Fe/H] = --1.37 (1$\sigma$ = 0.58 dex) while for LMC stars is
[Fe/H] = --1.49 (1$\sigma$ = 0.47 dex). As it is evident in the right
panel of Figure~\ref{fig:feh_histograms}, the metallicity distribution
of the likely SMC stars tends to be slightly more metal-rich than the
distribution of the LMC stars.


Our metallicity estimates are lower compared to the mean
(photometrically derived) metallicity using RR Lyrae stars reported by
\cite{Haschke12}, who found [Fe/H] = --1.5 and [Fe/H]=--1.70 for the
LMC and SMC, respectively. The difference of up to 0.3 dex may be
explained by a systematic difference between our method and the
light-curve based metallicity, or by the differences in the tracers
themselves. Another possible explanation is that this apparent
discrepancy could be due to a metallicity gradient as a function of
distance from the LMC/SMC centre. As shown in Fig.~4 of
\cite{Haschke12}, the mean metallicity of the RR Lyrae stars in the
LMC was derived based on the inner 8 kpc of the galaxy, while at
distances beyond 10 kpc, the metallicity drops by about 0.25 dex with
respect to the stars in the innermost region. Nonetheless, the
precision of our metallicity measurements is not good enough to
identify unambiguously the cause of this discrepancy.

\subsection{Comparison with simulations}\label{sec:simulations}

\begin{figure}
  \includegraphics[width=0.48\textwidth]{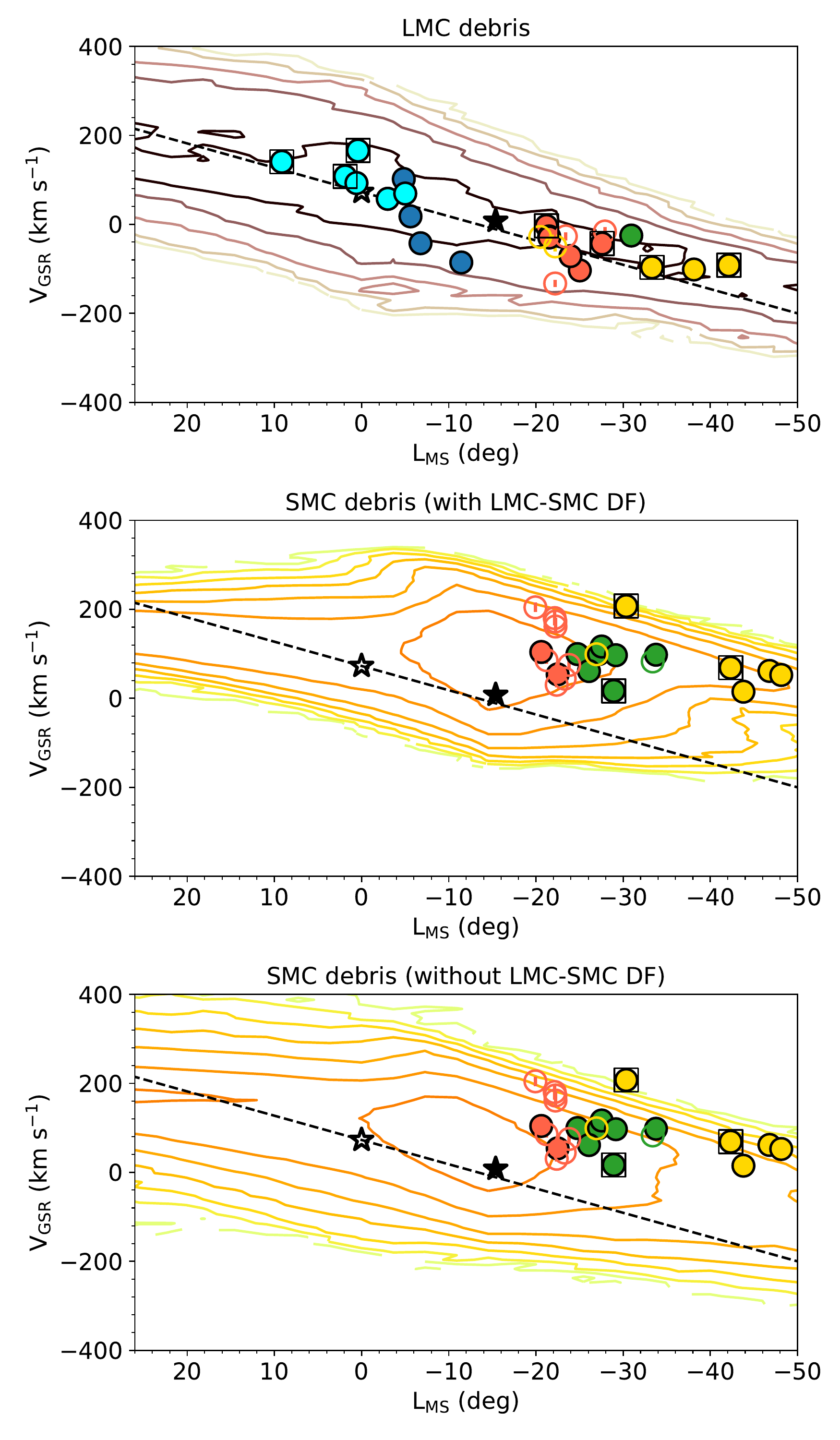}
  \caption{Phase-space distribution for LMC (top) and SMC 
    (middle and bottom) debris, based on the N-body simulations of
    \protect\cite{Jethwa16}. Middle and bottom panels show the results
    when the DF by the LMC on the SMC is included and
    when it is not, respectively. The top panel shows all our
    Magellanic BHB and BS stars that seem to follow the LMC's halo
    velocity trend (dashed line): i.e., all S1 stars and S2, S3 and
    S4 stars with V${}_{\rm GSR} <$~0~km~s${}^{-1}$, while middle and
    bottom panels show BHB and BS stars from S2, S3 and S4 with
    V${}_{\rm GSR} >$~0~km~s${}^{-1}$.}
    \label{fig:simulations}
\end{figure}

Figure~\ref{fig:simulations} compares the phase-space coordinates of
our Magellanic BHB and BS stars to the predictions from the
simulations of \cite{Jethwa16}. The underlying contours show the
density of the LMC and SMC debris particles, from a large suite of
simulations corresponding to a sample of 150 proper motion values for
the MCs, each containing 1000 tracer particles, for the following
galaxy masses: ($M_{\rm MW}$, $M_{\rm LMC}$, $M_{\rm SMC}$) = (75.0,
15.0, 2.0) $\times 10^{10}$ M${}_{\odot}$. We note
  that the comparison of our data with the numerical simulations
  should be taken as illustrative as we do not attempt to match our
  measurements to the mock debris distributions at hand. According to
  our preliminary analysis, while the phase-space distribution of the
  debris is clearly sensitive to the exact LMC/SMC mass ratio adopted
  as well as the orbital properties of the binary in-fall, some of the
  trends discussed here remain mostly unchanged. In the top panel,
the LMC debris are compared to the stars that appear to follow the
LMC's halo velocity gradient (see Figure~\ref{fig:lms_vgsr}): more
precisely, all S1 stars plus S2 to S4 stars with V${}_{\rm GSR} < 0$
km s${}^{-1}$. As discussed above, even stars located beyond 30 deg
from the LMC's center are located close to the predicted track of the
LMC's debris in phase-space. Moreover, it seems that the more
metal-poor S1 BHB stars are following a different path compared to the
bulk of the simulations, albeit well inside the region predicted to be
occupied by the LMC debris. The slight difference in the behaviour of
the S1 stars with different metallicities is curious, however. As
Figure~\ref{fig:lms_bms} demonstrates, the metal-richer S1 stars are
on average much closer to the LMC --many appear to coincide with the
disc spur described in \citet{Mackey16} and several are likely to be
still bound to the LMC. On the other hand, the metal-poorer members of
S1 all lie very far from the LMC, and in fact occupy a very different
range of $B_{\rm MS}$, namely $20^{\circ}<B_{\rm MS}<30^{\circ}$ as
opposed to $B_{\rm MS}<20^{\circ}$.

The BHB and BS stars with V${}_{\rm GSR} > 0$ km s${}^{-1}$ in our
sample appear to be somewhat at odds with the predicted phase-space
distribution of the LMC debris based on the \cite{Jethwa16}
simulations. These also do not follow the LMC velocity trend (see
Figure~\ref{fig:lms_vgsr}).  Given their mean velocity of $V_{\rm
  GSR}$ = 95 km s$^{-1}$, with a velocity dispersion of $\sim$ 50 km
s$^{-1}$, it is unlikely that the stars in this subgroup represent a
random sample of the MW halo. In fact, we speculate that these are
instead a part of the SMC debris also expected to litter this area of
the sky \citep[see e.g.,][]{Besla10, Olsen11, Belokurov17,
  Deason17}. Accordingly, middle and bottom panels of
Figure~\ref{fig:simulations} show the results from \cite{Jethwa16}
simulations for the SMC particles. The difference between these two
panels is the inclusion (exclusion) of the dynamical friction (DF)
exerted by the LMC onto the SMC: middle (bottom) panel shows the
prediction from the simulation when it is on (off). Note that in these
simulations, the DF of the MW on the MCs is always included. The group
of Magellanic BHB and BS stars which have V${}_{\rm GSR} > 0$ km
s${}^{-1}$ is overplotted, as well as the LMC's halo velocity
gradient (dashed line) for reference. There appears to be a better
match between these ``faster'' stars and the SMC debris if the LMC-SMC
DF is taken into account. Note that the effects of the DF, both from
the MW and from the LMC, is the main difference between the
simulations of \cite{Jethwa16} and those carried out by \cite{Besla10}
and \cite{Belokurov17}.


%

\begin{figure}
  \includegraphics[width=0.48\textwidth]{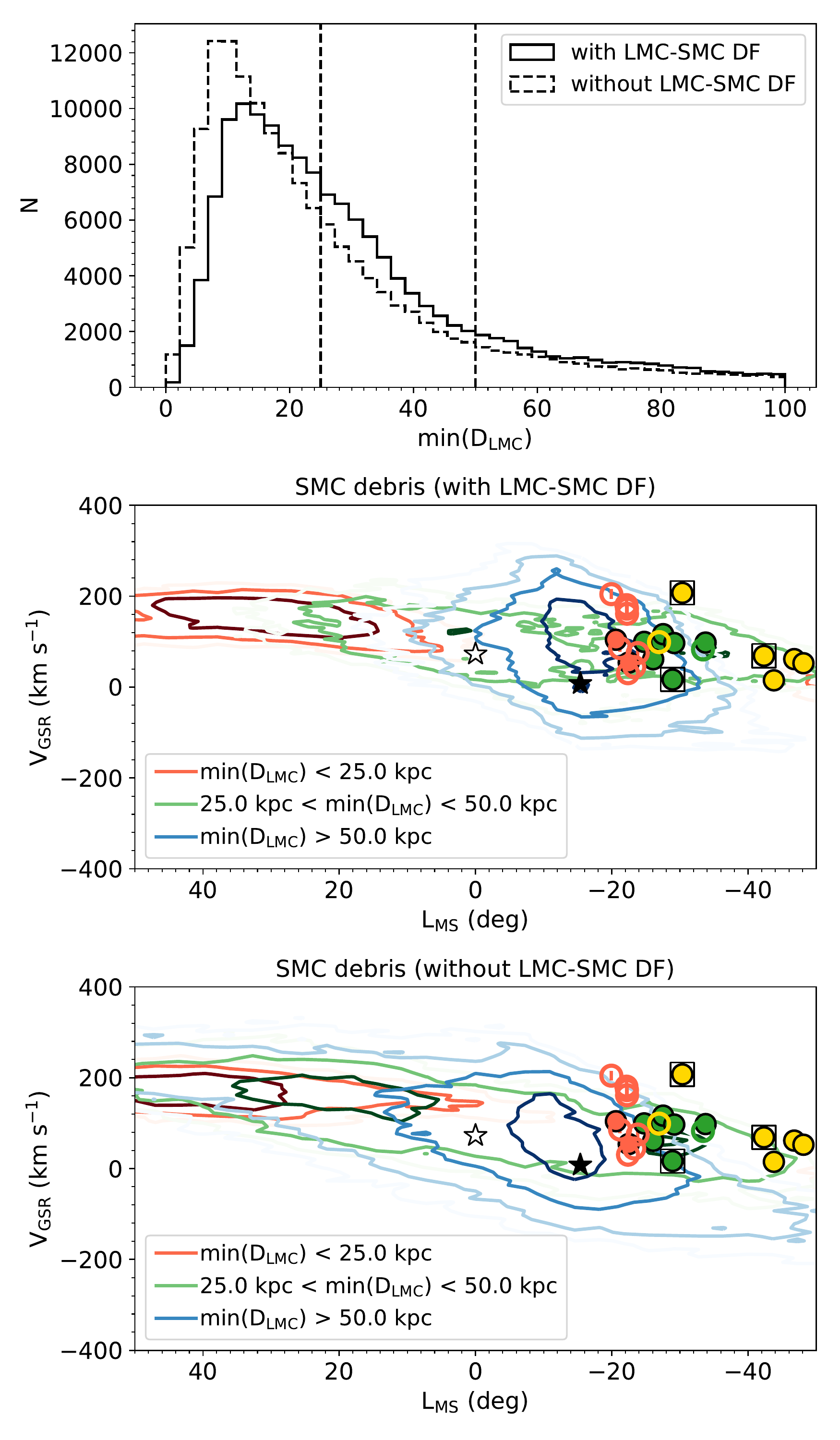}
  \caption{{\it Top panel}: Distribution of minimum distance to the
    LMC for 1000 particles, in 150 realizations of proper motion
    values from the simulations of \protect\cite{Jethwa16}, for those
    including and not including the DF effect of the LMC on the SMC
    debris particles (solid and dashed lines, respectively). {\it
      Middle and bottom panels}: contour density of SMC debris
    particles for simulations with and without LMC-SMC DF
    included. Red, green and blue contours correspond to particles
    which reach a minimum distance to the LMC, at a given point of
    their orbits, of less than 25 kpc, between 25 to 50 kpc and
    greater than 50 kpc, respectively.}
    \label{fig:mindistance}
\end{figure}

Interestingly, the comparison of the three panels of Figure
~\ref{fig:simulations} shows that at negative $L_{\rm MS}$, the SMC
debris encompasses a broader range of radial velocities as compared to
the LMC particles. In fact, the cloud of the SMC particles is
sufficiently broad to accommodate the stars with both negative and
positive $V_{\rm GSR}$. Superficially, this is unusual given that the
tidal stream width typically correlates with the progenitor's mass
\citep[][]{Erkal16a} and the SMC is at least 5-10 times lighter than
the LMC. We suspect that the phase-space distribution of the SMC
particle has been affected by the interaction with the LMC. To look
for signs of the LMC influence on the SMC debris, we split the SMC
particles into three groups according to the distance of their minimal
approach to the LMC, min$(D_{\rm LMC})$. The distribution of the
approach distances for all particles is shown in the top panel of
Figure~\ref{fig:mindistance}. A large number of the SMC particles in
the simulation have passed as close as $\lesssim$ 10 kpc (the peak of
the distribution) from the LMC. A considerable fraction of the
particles has reached a minimum distance between 25 and 50 kpc. There
remains, however, a significant number of particles that interacted
very weakly with the LMC, never approaching closer than 50 kpc.

The middle and bottom panels of Figure~\ref{fig:mindistance} show the
distributions of the three groups of SMC particles described above in
phase-space. Particles that have sustained the strongest perturbation
from the LMC, coming as close as 25 kpc to it, lie predominantly at
positive $L_{\rm MS}$ (red contours). Unsurprisingly, these are also
the closest to the LMC on the sky. At negative $L_{\rm MS}$, the
velocity distribution bifurcates into two sequences: one corresponding
to the stars with min$(D_{\rm LMC})>50$ kpc (blue), located at
negative radial velocities, and the one corresponding to the stars
with $25<$ min$(D_{\rm LMC}) [{\rm kpc}] <50$ (green) with positive
$V_{\rm GSR}$. Note that these two sequences are visible regardless of
whether the LMC-SMC DF is taken into account or not. Alternatively,
the min$(D_{\rm LMC})$ parameter can be considered as a proxy for the
``unbinding'' time --the epoch at which the SMC stars have been fully
stripped, i.e. stopped being influenced by their progenitor galaxy and
started orbiting the MW-LMC system.

%
%


\section{Discussion and Conclusions}\label{sec:dandc}

We have used FORS2 on the VLT to obtain medium-resolution spectroscopy
of 104 candidate A-type stars in the vicinity of the MCs. These were
selected to lie along the four candidate streams announced in
\citet{Be2016}. Of these, 25 objects turned out to be contaminants,
predominantly QSO, and WDs. The spectra of the remaining 79 stars show
prominent Balmer series absorption lines in the wavelength range of
3600 - 5110 \AA. We model the observed spectra with a combination of
S\'ersic profiles for the absorption and a fifth-degree polynomial for
the continuum (see Figure~\ref{fig:fit_example}). Using the shapes of
the Balmer lines we devise a classification scheme to separate the BHB
stars from BSs, which we test on a large sample of high quality
spectra of similar resolution, provided by the SDSS (see
Figure~\ref{fig:bhb_bs_sdss}).  According to our classification, 24
stars are of BHB type, 45 are BSs and 10 have uncertain classification
(see Figure~\ref{fig:cbparameters}). All of our BHB stars lie beyond
35 kpc from the Sun, and therefore we consider the 15 BS stars with
Galactocentric distances larger than 35 kpc as part of the final
sample of stars of likely Magellanic origin. Additionally, we use Ca
II K line at 3933 \AA\ to estimate the stellar metallicity. We devise
an empirical calibration based on the equivalent width values
available for a sample of BHBs and BSs with available SDSS
spectroscopic data (see Figures~\ref{fig:ca_teff_feh} and
\ref{fig:ca_feh_fitlines}).

Stars classified as BHBs and BSs show different behavior in the
phase-space spanned by the line-of-sight velocity $V_{\rm GSR}$ and
the MS longitude $L_{\rm MS}$ (Figure~\ref{fig:lms_vgsr}). Velocities
of the BS stars appear consistent with a Gaussian distribution
centered on $V_{\rm GSR}=0$ and having a dispersion of 88 km s$^{-1}$. This
can be contrasted with the distribution of the BHB stars, many of
which avoid $V_{\rm GSR}=0$ and instead line up in several narrow
sequences at positive and negative $V_{\rm GSR}$. Such obvious
difference in the phase-space distribution provides further support
for our BHB/BS classification.

Of the four candidate streams presented in \citet{Be2016}, only two,
S1 and S2, appear kinematically distinct. The S2 stream not only has a
relatively small velocity dispersion of 15.0 km s$^{-1}$ (excluding
the S2 star with $V_{\rm GSR}$ < 0), it also has narrow Ca K
equivalent width --and thus metallicity-- distribution (see
Figure~\ref{fig:ews_vgsr}). Curiously, the S1 stars with the highest
metallicity are also the ones closest to the LMC's disc (see
Figure~\ref{fig:lms_bms}). In fact, several of these appear to overlap
with the spur detected by \citet{Mackey16}. It is not therefore
impossible that rather than metal-rich BHBs (some of) these could
instead be bona-fide Young Main Sequence stars. The remaining,
metal-poor S1 stars follow a different phase-space track, reaching as
far as $\sim30^{\circ}$ north of the LMC's center. Therefore, we
conjecture that the S1 stars considered here actually sample two
different structures in the northern side of the LMC. Comparing the
velocity and the metallicity distributions of the stars in S1 and S2,
we conclude that S2 --metal poor and cold-- is the best candidate to
date for a tidal stream from a low-mass system accreted by the LMC.

Looking at all of the stars in our ``Magellanic'' sample, i.e. 24 BHBs, 15 BSs 
and 10 BHB/BSs, we note two clear trends of the radial velocity as a function of 
MS longitude. First, most of the S1 stars and a large fraction of S3 and S4 ones 
follow a clear velocity trend where $V_{\rm GSR}$ steadily decreases with 
angular distance from the LMC (see dashed line in Figure~\ref{fig:lms_vgsr}).  
Interestingly, it is impossible to distinguish between the stars still bound to 
the LMC and the stars in its tidal tail based on their $(V_{\rm GSR}, L_{\rm 
MS})$ position. This is because the projection of the LMC's space velocity onto 
the line of sight and the centroid of the tidal debris coincide in phase-space 
(see Figure~\ref{fig:simulations}). Regardless of whether these stars are still 
bound to or recently stripped from the LMC, these objects (marked with ``M1'' in 
Table~\ref{tab:targets}) are probably of Magellanic origin.

We also investigate whether any of the stars in our sample could have
originated in the SMC. According to Figure~\ref{fig:simulations}, all
of them could!  Even for lower (1.0 $\times 10^{10}
  M_{\odot}$) and higher (3.0, 4.0 $\times 10^{10} M_{\odot}$) SMC
  masses, the velocity trend is compatible with the group of stars at
  high GSR velocities located between $L_{\rm MS} = -20 \degr$ to
  $-50$\degr (P. Jethwa, private communication). Note that, for a
large number of stars, it is impossible to identify whether they were
(are) part of the LMC or the SMC based on the data in
hand. Nonetheless, there exists a group of stars with $V_{\rm GSR}>0$
and $L_{\rm MS}<-20^{\circ}$ with velocities significantly higher
compared to what is predicted for the LMC particles at a given MS
longitude (marked with ``M2'' in Table~\ref{tab:targets}). However,
the kinematics of these stars can be easily explained if they were
stripped from the SMC. As Figure~\ref{fig:mindistance} illustrates, at
negative $L_{\rm MS}$ there exists a bifurcation in the velocity
distribution of the SMC particles. Stars populate different branches
of the velocity bifurcation according to the distance of the closest
approach to the Large Cloud. This effect can plausibly be explained by
two phenomena. First, the minimal distance to the LMC likely
correlates with the time of unbinding of a particle from the
SMC. Thus, the velocity bifurcation is the natural consequence of the
orbital evolution of the particles in a combined potential of the
LMC+MW.  A second, related explanation is possible where a certain
distance exists within which the LMC influences the orbits of the SMC
debris enough to elevate them to a higher velocity.

Note that at large angular distances from the LMC, given the data in
hand, it is not possible to identify securely which stars were
stripped from the Clouds and which belong to the virialised MW
halo. Therefore, while we confirm the Magellanic origin for S1 and S2
stars, for association between the Clouds and S3 and S4 stars should
be considered as a speculation, which if proven can shed light onto
the history of interaction between the LMC and SMC and the Milky
Way. We look forward to testing this hypothesis using the astrometric
information from the Gaia satellite.

\section*{Acknowledgements}

Based on observations collected at the European Organisation for Astronomical 
Research in the Southern Hemisphere under ESO programme(s) 098.B-0454(A).

The research leading to these results has received funding from the
European Research Council under the European Union's Seventh Framework
Programme (FP/2007-2013)/ERC Grant Agreement no. 308024.

This project is supported by CONICYT's PCI program through grant DPI20140066. 
Additional support is provided by the Ministry for the Economy, Development, and 
Tourism's Iniciativa Cient\'ifica Milenio through grant IC\,120009, awarded to 
the Millennium Institute of Astrophysics; by Proyecto Fondecyt Regular 
\#1171273; and by Proyecto Basal PFB-06/2007. C.N. acknowledges support from 
CONICYT-PCHA grant Doctorado Nacional 2015-21151643. M.C. gratefully 
acknowledges the additional support provided by the Carnegie Observatories 
through its
Distinguished Scientific Visitor program. JAC-B acknowledges financial support 
to CONICYT-Chile FONDECYT Postdoctoral Fellowship 3160502 and CAS-CONICYT 17003. 
S.~D. acknowledges support from Comit\'e Mixto ESO-GOBIERNO DE CHILE.




\bibliographystyle{mnras}
 \bibliography{references}

\bsp	
\label{lastpage}

\twocolumn
\end{document}